\documentclass[12pt]{iopart}

\usepackage[utf8]{inputenc}
\usepackage[T1]{fontenc}
\usepackage[english]{babel}

\usepackage[pdftex]{graphicx}
\usepackage{bbm,amsthm,amsfonts,amssymb,amsopn,amsmath}
\usepackage{dsfont}
\usepackage{layouts,enumerate}
\usepackage{mathtools,physics}
\usepackage{url,hyperref,cite}
\usepackage{pgf,tikz,hhline}
\usepackage{iopams}
\usepackage{calc,xstring,ifthen}
\usepackage{pdftexcmds,adjustbox}
\usetikzlibrary{external,arrows,calc,matrix,backgrounds,fit,positioning,
decorations.pathmorphing,decorations.markings}
\usepackage{bbm,bm}
\usepackage{cleveref}
\usepackage{caption,subcaption,float}

\makeatletter
\renewcommand*\env@matrix[1][\arraystretch]{%
  \edef\arraystretch{#1}%
  \hskip -\arraycolsep
  \let\@ifnextchar\new@ifnextchar
  \array{*\c@MaxMatrixCols c}}
\makeatother
\renewcommand{\tr}{\mathrm{tr}}

\newcommand{\re}{\mathrm{e}}
\newcommand{\hre}{\hat{\mathrm{e}}}
\newcommand{\oo}[1]{{[ #1 ]}}
\renewcommand{\vec}[1]{{\mathbf{#1}}}
\newcommand{\unit}{\mathbbm{1}}
\newcommand{\Ue}[1]{U^{#1\,(\mathrm{e})}}
\newcommand{\Uo}[1]{U^{#1\,(\mathrm{o})}}
\newcommand{\vV}{\vec{W}^{\prime}}
\newcommand{\vW}{\vec{W}^{\phantom{\prime}}}
\newcommand{\vY}{\vec{X}^{\prime}}
\newcommand{\vX}{\vec{X}^{\phantom{\prime}}}
\newcommand{\V}{{W}^{\prime}}
\newcommand{\W}{{W}^{\phantom{\prime}}}
\newcommand{\Y}{{X}^{\prime}}
\newcommand{\X}{{X}^{\phantom{\prime}}}

\newcommand{\be}{\begin{eqnarray}}
\newcommand{\ee}{\end{eqnarray}}
\newcommand{\bea}{\begin{eqnarray}}
\newcommand{\eea}{\end{eqnarray}}
\newcommand{\ul}[1]{\underline{#1}}

\definecolor{sp1}{RGB}{110,184,214}
\definecolor{sp2}{RGB}{179,209,162}
\definecolor{border}{rgb}{0.3,0.3,0.3}
\definecolor{ruleText}{rgb}{0.1,0.1,0.1}

\def\a{0.4}

\newcommand\crectangle[5]{
  \ifthenelse{\equal{#3}{1}}{\colorlet{tempcolor}{sp1}}{\ifthenelse{\equal{#3}{2}}{\colorlet{tempcolor}{sp2}}{\colorlet{tempcolor}{white}}}
  \draw[tempcolor,fill=tempcolor] ({\a*(#1)},{\a*(#2-1)})  -- ({\a*(#1+1)},{\a*(#2)})  -- ({\a*(#1)},{\a*(#2+1)})  
  -- ({\a*(#1-1)},{\a*(#2)})  -- cycle;
  \draw[#4,thick] ({\a*(#1)},{\a*(#2-1)})  -- ({\a*(#1+1)},{\a*(#2)})  -- ({\a*(#1)},{\a*(#2+1)});
  \draw[#5,thick] ({\a*(#1)},{\a*(#2-1)})  -- ({\a*(#1-1)},{\a*(#2)})  -- ({\a*(#1)},{\a*(#2+1)});
}

\newcommand\dashedcrectangle[5]{
  \ifthenelse{\equal{#3}{1}}{\colorlet{tempcolor}{sp1}}{\ifthenelse{\equal{#3}{2}}{\colorlet{tempcolor}{sp2}}{\colorlet{tempcolor}{white}}}
  \draw[tempcolor,fill=tempcolor] ({\a*(#1)},{\a*(#2-1)})  -- ({\a*(#1+1)},{\a*(#2)})  -- ({\a*(#1)},{\a*(#2+1)})  
  -- ({\a*(#1-1)},{\a*(#2)})  -- cycle;
  \draw[#4,dashed] ({\a*(#1)},{\a*(#2-1)})  -- ({\a*(#1+1)},{\a*(#2)})  -- ({\a*(#1)},{\a*(#2+1)});
  \draw[#5,dashed] ({\a*(#1)},{\a*(#2-1)})  -- ({\a*(#1-1)},{\a*(#2)})  -- ({\a*(#1)},{\a*(#2+1)});
}

\begin{document}
\title[Two-species reversible cellular automata]
{On two reversible cellular automata with two particle species}

\author{Katja Klobas$^{1}$, Toma\v{z} Prosen$^2$}
\address{$^1$ Rudolf Peierls Centre for Theoretical Physics,
Oxford University, Parks Road, Oxford OX1 3PU, United Kingdom}
\address{$^2$ Department of Physics, Faculty of Mathematics and Physics,
University of Ljubljana, Jadranska 19, 1000 Ljubljana, Slovenia}
\ead{katja.klobas@physics.ox.ac.uk} %, tomaz.prosen@fmf.uni-lj.si}

\date{\today} % date

\begin{abstract} % abstract
 We introduce a pair of time-reversible models defined on the discrete
    space-time lattice with $3$ states per site, specifically, a vacancy and a
    particle of two flavours (species). The local update rules reproduce the
    Rule 54 reversible cellular automaton when only a single species of
    particles is present, and satisfy the requirements of flavour exchange
    ($C$), space-reversal ($P$), and time-reversal ($T$) symmetries. We find
    closed-form expressions for three local conserved charges and provide an
    explicit matrix product form of the grand canonical Gibbs states, which are
    identical for both models. For one of the models this family of Gibbs
    states seems to be a complete characterisation of equilibrium (i.e.\ space
    and time translation invariant) states, while for the other model we
    empirically find a sequence of local conserved charges, one for each
    support size larger than $2$, hinting to its algebraic integrability. Finally, we
    numerically investigate the behaviour of spatio-temporal correlation
    functions of charge densities, and test
    the hydrodynamic prediction
    for the model with exactly three local charges.
    Surprisingly, the numerically observed `sound velocity' does not match the
    hydrodynamic value.
    The deviations are either significant, or they decay extremely
    slowly with the simulation time, which leaves us with an open question for
    the mechanism of such a glassy behaviour in a deterministic locally
    interacting system.
\end{abstract}
% \keywords{ <sample text> } % keywords
%\submitto{jpa}
\section{Introduction}
Understanding dynamical phenomena in large systems of interacting particles is
one of the central issues of statistical mechanics~\cite{spohn2012large}.
Although the microscopic laws of motion in physical systems are \emph{deterministic},
one often simplifies the models by considering a stochastic
microscopic description to obtain analytically tractable models; such
as simple exclusion processes and reaction-diffusion processes~\cite{schutz2001}.
Nevertheless, the ultimate goal %of statistical physics
should be to derive the emergence of universal macroscopic statistical laws,
say the Fick's or Fourier's law of diffusive transport and precise conditions
for various super- or sub-diffusive anomalies, from deterministic reversible
laws of motion. Ideally, one would like to achieve this in typical
Hamiltonian systems of interacting particles, such as Fermi-Pasta-Ulam-Tsingou (FPUT)
chains~\cite{Lepri03,Lepri16}, but at the moment this seems out of reach
within any rigorous framework. Arguably the most suitable systems for this
purpose, which can be viewed as caricatures of Hamiltonian dynamics containing
its essential features, are reversible cellular automata (RCA).

The simplest type of RCA, namely in one spatial dimension and with a 3-site Margolus
neighbourhood, have been completely classified in Ref.~\cite{bobenko1993two},
where it has been pointed out that the reversible cellular automaton with the
rule code 54 (RCA54)\footnote{Equivalent to rule code 250R of
Ref.~\cite{takesue1987reversible}, which used a slightly more complicated
setup.} has all the features of an integrable interacting particle system with
solitonic excitations and conserved currents. Indeed, in recent years,
equilibrium and non-equilibrium statistical mechanics of RCA54 has been
essentially completely solved (see Ref.~\cite{RCA54review} for a review), where
probably the main achievement was a rigorous derivation of diffusive dynamical
structure factor~\cite{klobas2019timedependent}. The main reason for the
remarkable utility of RCA54 lies in the underlying algebraic structures which
seem to go beyond Yang-Baxter integrability and allow for exact access of
time-dependent statistical quantitites, such as dynamical correlation functions
of local observables. Nevertheless, the observations of anomalous
(superdiffusive) transport in some Hamiltonian particle systems, like FPUT and
related anharmonic oscillator chains \cite{Lepri03,Lepri16}, or integrable
chains with nonabelian
symmetries~\cite{ljubotina2019kardar,krajnik2020integrable,bulchandani2021superdiffusion},
possess a natural question about the existence of an exactly solvable RCA featuring anomalous
transport properties. The obvious playground to look for such models are
multispecies RCA which would reduce to RCA54 under some limiting situations,
say for dynamics restricted to configurations with single particle species or
isolated solitonic excitations. Permutation symmetry among the species (or
particle flavours) may then serve as a kind of discrete analog of $\mathrm{SU}(2)$
symmetry which was the minimal requirement for observing anomalous spin
transport of Kardar-Parisi-Zhang universality in integrable spin
chains~\cite{ljubotina2019kardar,krajnik2019kadar}.

In this paper we construct two one-dimensional RCA with three states per
site, which naturally generalise RCA54 to two particle species. We show that,
by requiring the update rules to be invariant under the species permutation
(conjugation), space reversal and time reversal, the model is fixed up to one
binary choice. One choice corresponds to the particle flavour conservation in
the annihilation-creation process. For this model we find a numerical evidence
of an extended number of conservation laws with local charge density, which hints
to its integrability, but the Yang-Baxter integrability structures are still not
known. By flipping the species conjugation in the particle
annihilation/creation processes, we obtain the second model, which has exactly
the same local charges of support size 2, but lacks additional local conservation
laws and is hence likely to be non-integrable.  We provide a simple algebraic
characterization of Gibbs equilibrium states for both models, and extend the
result to a stationary generalized Gibbs state with one higher conservation law
for the integrable model. Finally, we simulate spatio-temporal correlation
functions of the three local charge densities. In the integrable case the situation
seems complex and difficult to describe quantitatively due to the coupling
among an extensive number of charges, while the non-integrable case with
exactly three charges may be amenable to nonlinear fluctuating hydrodynamics
which has been successfully applied to anharmonic (e.g.\ FPUT) particle
chains~\cite{spohn2014nonlinear}.
We find a small but potentially
significant deviation from the hydrodynamic prediction even for the sound
velocity, which is determined by the standard Euler-scale hydrodynamics.
The results are not inconsistent with the numerically determined 
velocity eventually converging to the value predicted by hydrodynamics at 
`astronomically' long times, but the reason for existence of such long time
scales in such a simple parameter free model remains obscure.

%significant deviation from the theoretical prediction of NFH even for the
%simplest quantity, the sound velocity. The results are not inconsistent with
%eventual convergence of the simulated sound velocity to NFH value at
%`astronomically' long times, but the reason for existence of such long time
%scales in such a simple parameter free model remains obscure.

\section{Definition of the dynamics}
\subsection{Time evolution of configurations}
\begin{figure}
  \centering
  \begin{minipage}{.5\textwidth}
    \centering
      \includegraphics[width=\textwidth]{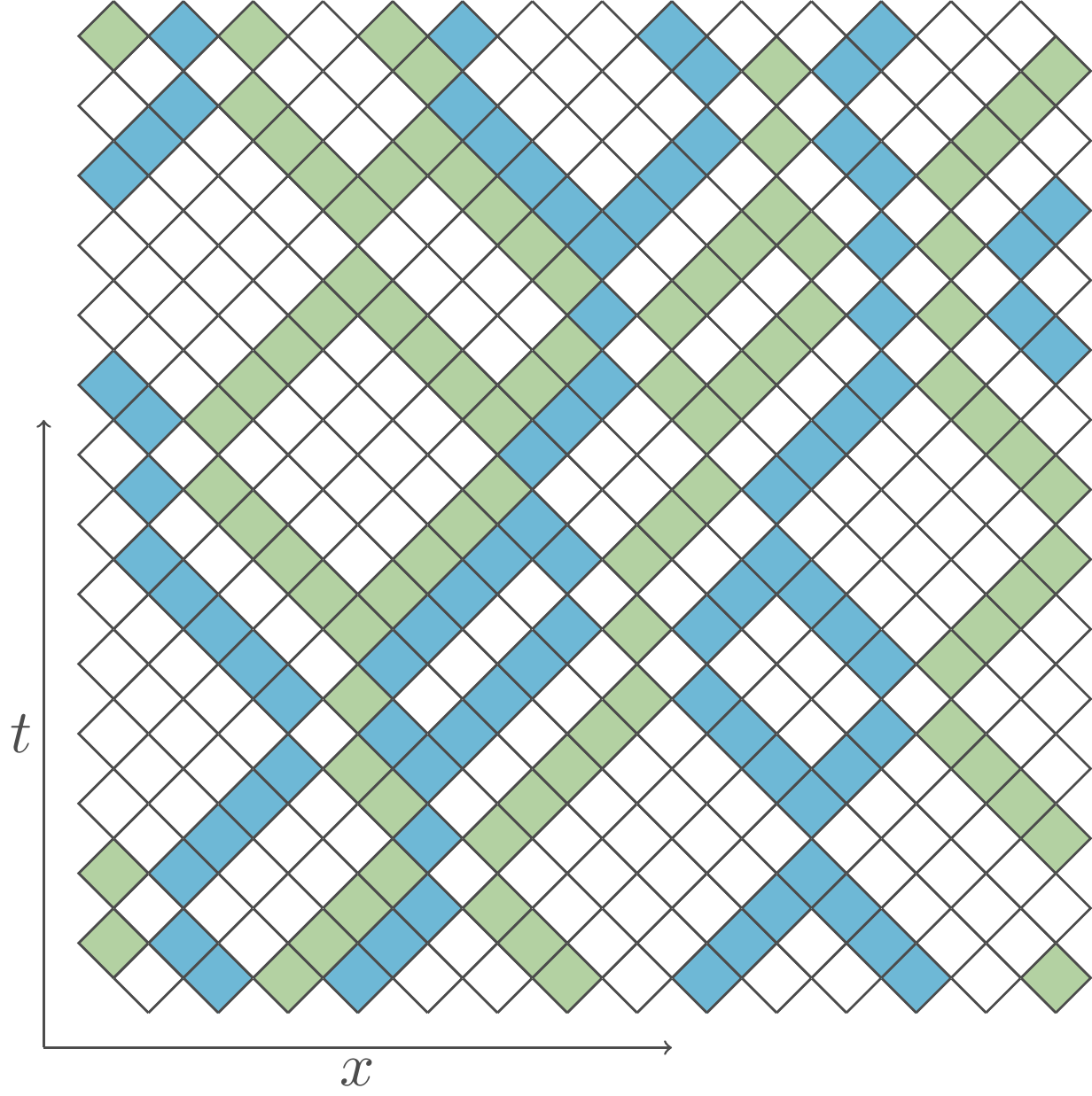}
  \end{minipage}%
  \begin{minipage}{.5\textwidth}
    \centering
      \includegraphics[width=\textwidth]{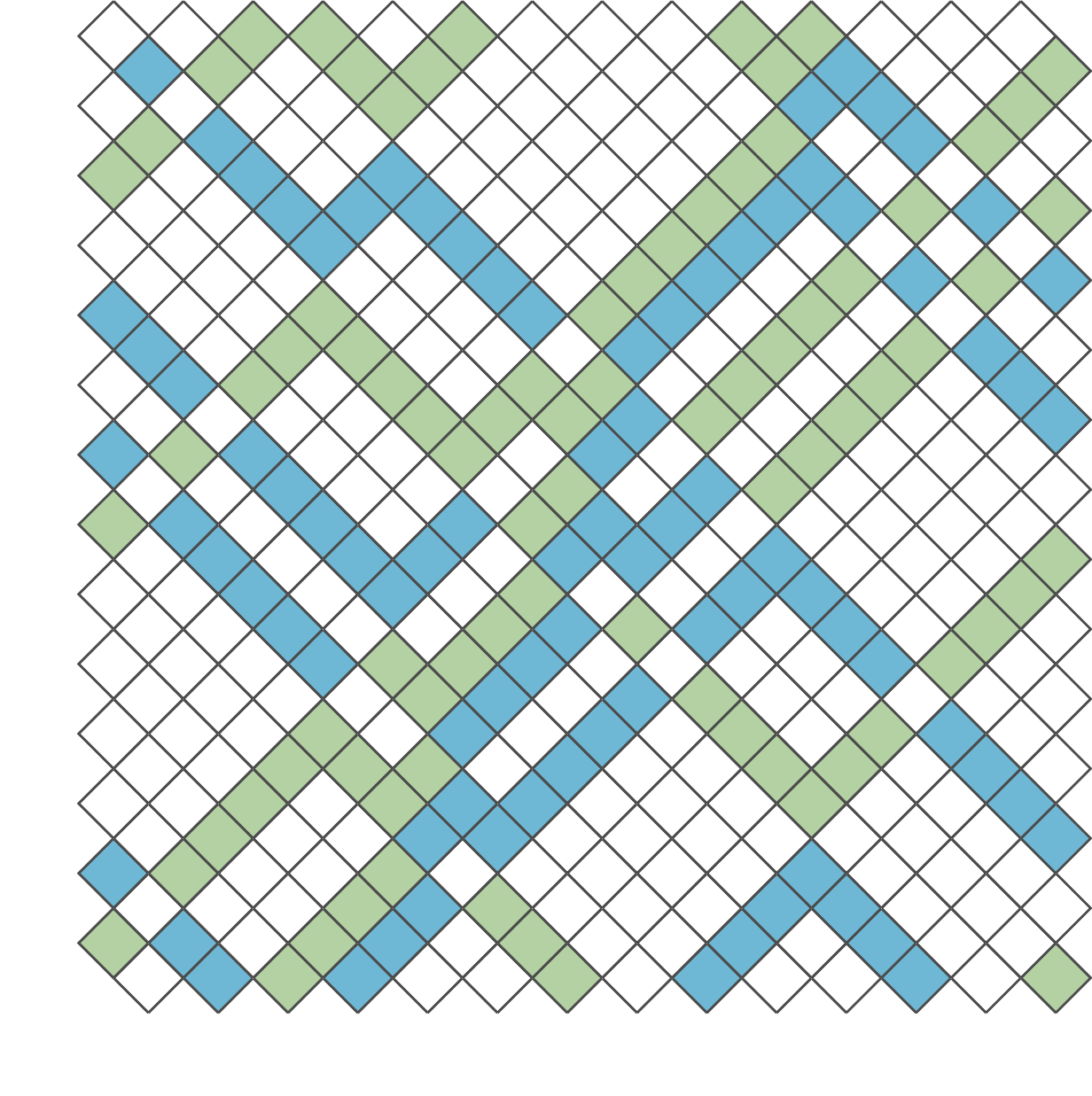}
  \end{minipage}
  \caption{\label{fig:dynExamples} Example of dynamics induced by time-evolution rules.
  The initial configuration (at the bottom) is evolved in time using the maps
  $\chi_{1}$ on the left, and $\chi_2$ on the right.}
\end{figure}

The main idea behind the construction of the dynamical rules is an attempt to
generalize the dynamics of RCA54 to a model with two particle species. It is
defined on the lattice of length~$2n$, where each site can be either empty
(denoted by $0$) or occupied by a particle of one of two species ($1$ and~$2$).
The configuration at time~$t$ is given by a ternary
string~$\underline{s}^t=(s^t_1,s^t_2,\ldots,s^t_{2n})$, $s^t_j\in\{0,1,2\}$,
and time evolution is defined in two time-steps,
\begin{eqnarray}
  \underline{s}^{t+1}=
  \begin{cases}
    \chi_{\alpha}^{\mathrm{e}}(\underline{s}^t),\qquad & t\equiv 0\pmod{2},\\
    \chi_{\alpha}^{\mathrm{o}}(\underline{s}^t), &  t\equiv 1\pmod{2},
  \end{cases}
\end{eqnarray}
where the subscript~$\alpha\in\{1,2\}$ discriminates between the two different
dynamical maps. At even time-steps the even sites change, while in the odd
time-steps the values at the odd sites change, 
\begin{eqnarray}
  \eqalign{
    \chi_{\alpha}^{\mathrm{e}}(s_1,s_2,\ldots,s_{2n})
    =(s_1,s_2^{\prime},s_3,s_4^{\prime},\ldots,s_{2n}^{\prime}),\\
    \chi_{\alpha}^{\mathrm{o}}(s_1,s_2,\ldots,s_{2n})
    =(s_1^{\prime},s_2,s_3^{\prime},s_4,\ldots,s_{2n}),
  }
\end{eqnarray}
where the updated values~$s_j^{\prime}$ are given by a local three-site update
rule
\begin{eqnarray}
  s_j^{\prime}=\chi_{\alpha}(s_{j-1},s_j,s_{j+1}).
\end{eqnarray}
Local maps~$\chi_1$ and~$\chi_2$ differ only in one three-site configuration,
\begin{eqnarray}\label{eq:defDifferent}
  \chi_1(0,s,0)=s,\qquad \chi_2(0,s,0)=\overline{s},
\end{eqnarray}
where~$s\in\{1,2\}$ and $\overline{\cdot}$ denotes the flip (exchange) of particle species,
$\overline{s}\equiv (3-s)\pmod{3}$. All other rules coincide. To determine them,
we first require that they reproduce the RCA54 behaviour
when only one particle is present, which implies the following 
\begin{eqnarray}\label{eq:defSame1}
  \fl
  \eqalign{
    \chi_{\alpha}(0,0,0)=0,\qquad
    \chi_{\alpha}(0,0,s)=s,\qquad
    \chi_{\alpha}(0,s,s)=0,\qquad
    \chi_{\alpha}(s,0,0)=s,\\
    \chi_{\alpha}(s,0,s)=s,\qquad
    \chi_{\alpha}(s,s,0)=0,\qquad
    \chi_{\alpha}(s,s,s)=0.
  } \label{R54}
\end{eqnarray}
Furthermore, the local map $\chi_\alpha$ is required to be reversible, symmetric with respect to the
left-right reflection, and invariant under the flip of particle species\footnote{We can argue that our dynamical laws have, respectively, $T$, $P$, and $C$, symmetries.} $1\leftrightarrow 2$,
\begin{eqnarray}
  \eqalign{
    \chi_{\alpha}(s_1,\chi_{\alpha}(s_1,s_2,s_3),s_3)=s_2,\\
    \chi_{\alpha}(s_1,s_2,s_3)=\chi_{\alpha}(s_3,s_2,s_1),\qquad
    \chi_{\alpha}(\overline{s_1},\overline{s_2},\overline{s_3})=
    \overline{\chi_{\alpha}(s_1,s_2,s_3)},
  }
\end{eqnarray}
which, together with Eqs.~\eqref{R54}, fix the following update rules,
\begin{eqnarray}\label{eq:defSame2}\fl
  \chi_{\alpha}(0,s,\overline{s})=s,\qquad
  \chi_{\alpha}(s,\overline{s},s)=\overline{s},\qquad
  \chi_{\alpha}(s,\overline{s},0)=\overline{s},\qquad
  \chi_{\alpha}(s,0,\overline{s})=0.
\end{eqnarray}
Additionally, the remaining four transitions are determined by
\begin{eqnarray}\label{eq:defSame3}
  \chi_{\alpha}(s,s,\overline{s})
  =\overline{s},\qquad\chi_{\alpha}(s,\overline{s},\overline{s})=s,
\end{eqnarray}
resulting in stripes of alternating flavour configurations propagating freely
(see Figure~\ref{fig:dynExamples} and the discussion at the end of the
subsection).  Thus the time evolution is completely determined by
equations~\eqref{eq:defDifferent}, \eqref{eq:defSame1}, \eqref{eq:defSame2},
and~\eqref{eq:defSame3}.

\begin{figure}
  \centering
    \includegraphics[width=0.8\textwidth]{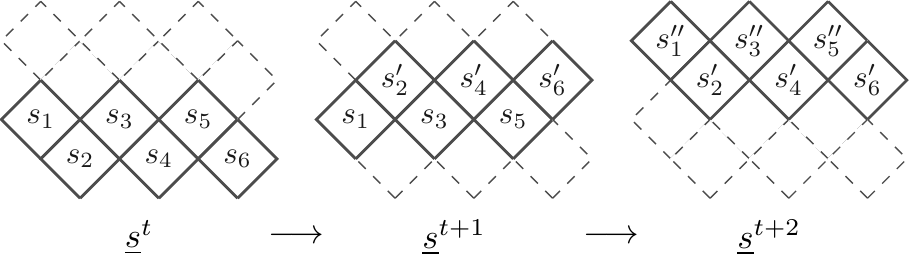}
  \caption{\label{fig:geometryOfTE}
  Graphical representation of the geometry of time evolution. In the first
  time-step the sites at the bottom of the zig-zag saw-shaped chain are updated according to the
  local time-evolution rules,
  $s_j^{\prime}=\chi_{\alpha}(s_{j-1},s_j,s_{j+1})$. This can be graphically
  expressed by removing the site with the old value~$s_j$ and add the updated
  value~$s_j^{\prime}$ to the top of the chain, so that the sites that were
  previously on the top are now lying on the bottom of the updated chain. In
  the second time step, the procedure is repeated,
  $s_j^{\prime\prime}=\chi_{\alpha}(s_{j-1}^{\prime},s_{j},s_{j+1}^{\prime})$,
  which completes the full period of time evolution.
  }
\end{figure}

\begin{figure}
  \centering
    \includegraphics[width=\textwidth]{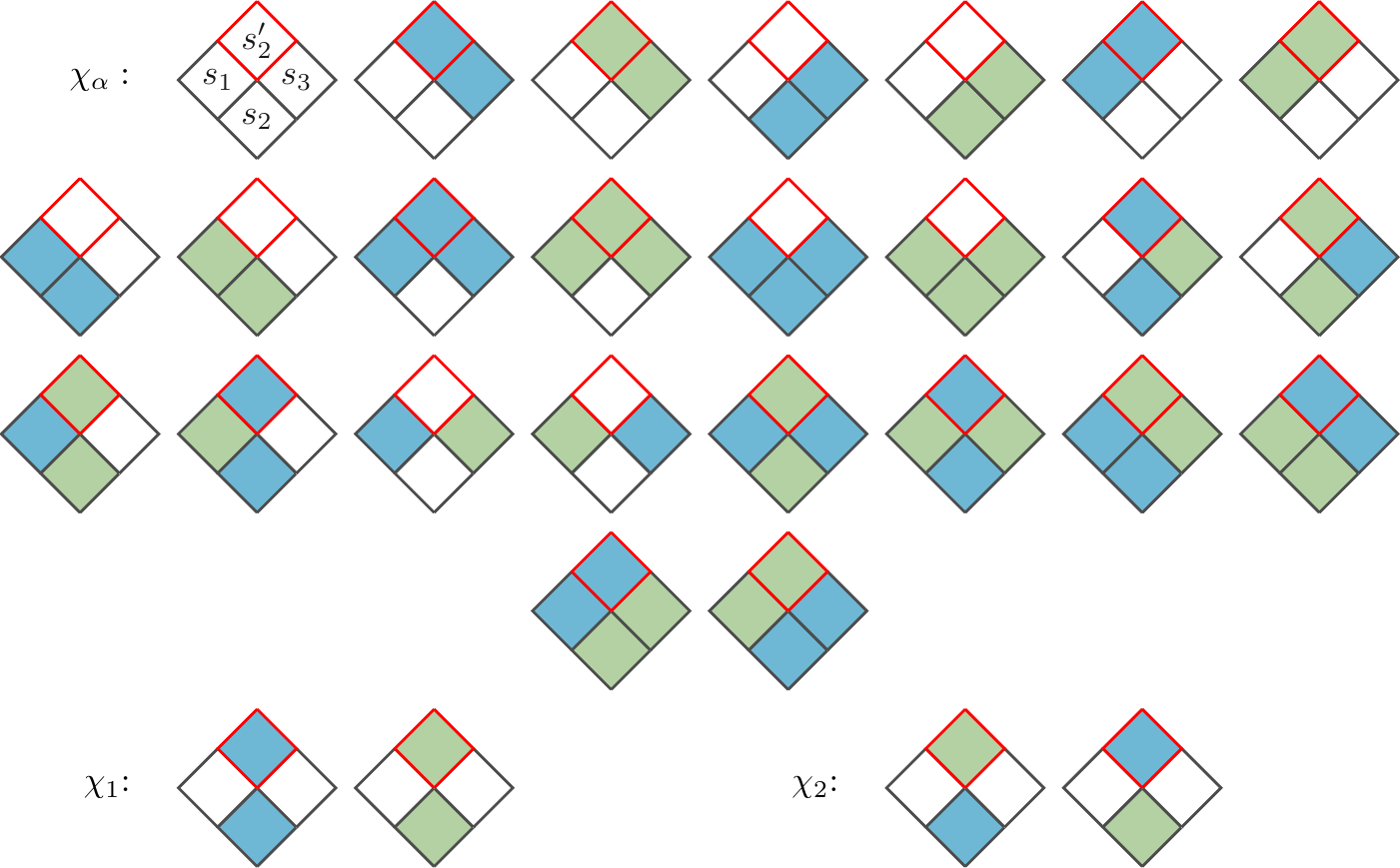}
  \caption{\label{fig:graphRules}
  Graphical representation of local time-evolution maps~$\chi_1$ and $\chi_2$.
  The red-bordered rectangle at the top represents the new central value
  $s_2^{\prime}$, determined as $s_2^{\prime}=\chi_{\alpha}(s_1,s_2,s_3)$.
  Blue and green coloured squares correspond to $s_j=1$ and $s_j=2$
  respectively, while white squares denote empty sites.  Time evolution in the
  two automata differs only in the last two diagrams,
  $\chi_1(0,s,0)\neq \chi_2(0,s,0)$, $s\in\{1,2\}$ (the bottom most row),
  while all the others are the same.
  }
\end{figure}

To visualize the dynamics the lattice can be drawn in a staggered zig-zag shape,
where at each time the bottom of the chain corresponds to the sites that are
being updated, while the top sites do not change, as is schematically shown in
Figure~\ref{fig:geometryOfTE}. Using this convention, the local time-evolution
rules can be summarized by a collection of~$27=3^3$ diagrams connecting the distinct three-site configurations at the bottom of $2\times 2$ rhombic plaquettes with the updated
site at the top, where the two particle flavours are represented by boxes of two distinct colours
and vacant sites by empty boxes, as shown in Figure~\ref{fig:graphRules}. The collection of $27$ update rule diagrams, as a subset of all $81=3^4$ rhombus configurations, is closed under separate actions of $C$ (species flip), $P$ (horizontal reflection), and $T$ (vertical reflection).  

\begin{figure}
  \centering
  \includegraphics[width=\textwidth]{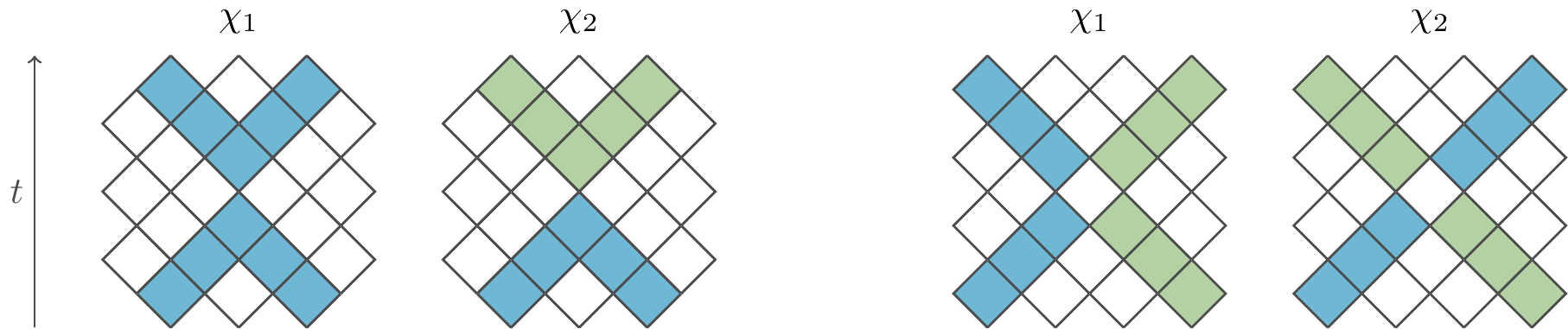}
  \caption{\label{fig:collExamples} Scattering of simple particles. To understand
  the interactions between different types of solitons, we start with a configuration
  with two oppositely moving particles (at the bottom) and evolve it in time (upwards).
  When the two particles are of the same type, they merge into one site and in the
  next time step reappear, and effectively obtaining a delay of~$1$ site. For the
  model given by~$\chi_1$ the particles emerging after the scattering are of the same
  type as before, while in the case of~$\chi_2$ the particle type is changed.
  If two particles of opposite colours meet, they do not obtain any delay and
  they either change their directions ($\chi_1$) or pass through each
  other undisturbed ($\chi_2$).
  }
\end{figure}

In both models, analogously to the RCA54, a pair of consecutive sites
with the same colour on the empty background behaves as a soliton that moves
with velocity $1$ either to the left or to the right. When two oppositely
moving particles of the same colour meet they temporarily merge into one
site and then reappear in the next time step, and thus effectively get
delayed for one site. However, as is shown on the left part of
Figure~\ref{fig:collExamples}, in one case (the automaton given by~$\chi_1$)
the particles preserve the colour, while in the model given by~$\chi_2$ the
colour of the particles gets exchanged when scattering. Similarly, the collision
of the particles of opposite types differs in the two models: in one case the
particles bounce off each other, while in the second system they pass through
each other unperturbed.  Another feature of the dynamics is the existence of
longer composite quasiparticles, which consist of stripes of occupied sites of
alternating colour moving at speed $1$ or $-1$. However, these excitations are
not stable under scattering, as can be seen in an example of the
time evolution starting from some typical (random) initial state in
Figure~\ref{fig:dynExamples}.

\subsection{States and observables}
Before continuing with the discussion of dynamical properties, let us first
define the notion of statistical states and observables, and introduce notation
used throughout the paper. 

A \emph{statistical} (also \emph{macroscopic}) state is a probability distribution
over the set of configurations and is completely determined by a $3^{2n}$ dimensional 
normalized vector with nonnegative components,
\begin{eqnarray}
  \vec{p}=\begin{bmatrix} p_0 & p_1 & \cdots p_{3^{2n}-1}\end{bmatrix}^T,
    \qquad p_s\ge0,\qquad \sum_{s=0}^{3^{2n}-1}p_s=1.
\end{eqnarray}
Each coefficient $p_s$ corresponds to the probability of the configuration
given by the ternary representation of~$s$,
\begin{eqnarray}
  s\equiv \underline{s}=(s_1,s_2,s_3,\ldots, s_{2n}),\qquad
  s=\sum_{j=1}^{2n} 3^{2n-j} s_j.
\end{eqnarray}
Time evolution of macroscopic states is given in terms of a~$27\times 27$ local propagator with the following matrix elements,
\begin{eqnarray}
  U^{\alpha}_{(s_1^{\prime},s_2^{\prime},s_3^{\prime}),(s_1,s_2,s_3)}
  = \delta_{s_1^{\prime},s_1}
  \delta_{s_2^{\prime},\chi_{\alpha}(s_1,s_2,s_3)}
  \delta_{s_3^{\prime},s_3},
\end{eqnarray}
where~$\alpha$ encodes the choice of the automaton. Time-evolution is given
by two distinct operators~$\Ue{\alpha}$ and $\Uo{\alpha}$,
\begin{eqnarray}
  \Ue{\alpha}=U^{\alpha}_{123}U^{\alpha}_{345}\cdots
  U^{\alpha}_{2n-, 2n, 1},\qquad
  \Uo{\alpha}=U^{\alpha}_{234}U^{\alpha}_{456}\cdots
  U^{\alpha}_{2n, 1, 2},
\end{eqnarray}
where~$U^{\alpha}_{j-1,j,j+1}$ is a three-site operator that nontrivially
acts on sites~$(j-1,j,j+1)$,
\begin{eqnarray}
  U^{\alpha}_{j-1, j, j+1} = \unit^{\otimes j-2} \otimes U^{\alpha} \otimes
  \unit^{2n-j-1},\qquad
  \unit=
  \begin{bmatrix}
    1& & \\ &1& \\ & &1\\
  \end{bmatrix}.
\end{eqnarray}
Time-evolution of the state~$\vec{p}$ can be therefore succinctly expressed
as
\begin{eqnarray}
  \vec{p}^{t+1} = \begin{cases}
    \Ue{\alpha} \vec{p}^{t},& t\equiv 0 \pmod{2},\\
    \Uo{\alpha} \vec{p}^{t},& t\equiv 1 \pmod{2}.
  \end{cases}
\end{eqnarray}

\emph{Observables} are real valued functions over the
set of configurations, $a: \mathbb{Z}_3^{2n} \to \mathbb{R}$,
that prescribe a unique value~$a(\underline{s})\in\mathbb{R}$
to each configuration $\underline{s}$. Expectation value of an
observable~$a$ in a state~$\vec{p}$ is given by
\begin{eqnarray}\label{eq:expVal}
  \expval*{a}_{\vec{p}}=\sum_{s=0}^{3^{2n}-1} a(s) p_s = \vec{a}\cdot \vec{p},\qquad
  \vec{a}=
  \begin{bmatrix}
    a(0)& a(1)& \cdots a(3^{2n}-1)
  \end{bmatrix},
\end{eqnarray}
where we identify~$a(s)\equiv a(\underline{s})$, with~$\underline{s}$ being
the ternary representation of~$s$. The last equality implies that the space
of observables can be understood as the vector space that is \emph{dual} to the
space of macroscopic states. Additionally, component-wise multiplication is well
defined,
\begin{eqnarray}
  (a\cdot b) (\underline{s}) = a(\underline{s})\,b(\underline{s}),
\end{eqnarray}
which makes the space of observables a commutative algebra.
\emph{Local} observables act nontrivially only on a finite subsection of the chain,
and their space is spanned by the following convenient basis,
\begin{eqnarray}
  \oo{q}_j(\underline{s})=\delta_{s_j,q},\qquad q\in\{0,1,2\}.
\end{eqnarray}
The one-site basis element~$\oo{q}_j$ is occupation indicator (also referred to as \emph{density}) of particles of flavour $q$
(or density of \emph{vacant sites} if~$q=0$) at the position~$j$. Any local observable
can be written as a linear combination of products of one-site observables. For
convenience we introduce the following short-hand notation for a complete basis of observables acting
on~$r$ consecutive sites,
\begin{eqnarray}
  \oo{q_1 q_2 q_3\ldots q_r}_j = 
  \oo{q_1}_j \cdot \oo{q_2}_{j+1} \cdots \oo{q_r}_{j+r-1}.
\end{eqnarray}

Time-evolution of observables is defined via the relation~\eqref{eq:expVal},
\begin{eqnarray}
  \expval*{a}_{\vec{p}^t} = \vec{a}\cdot \vec{p}^{t} = \vec{a}^t \cdot \vec{p}=
  \expval*{a^t}_{\vec{p}},
\end{eqnarray}
and is given in terms of the same time-evolution operators~$\Ue{\alpha}$,
$\Uo{\alpha}$,
\begin{eqnarray}
  \vec{a}^{2t}=\left(\Ue{\alpha}\Uo{\alpha}\right)^t \vec{a},\qquad
  \vec{a}^{2t+1}=
  \left(\Ue{\alpha}\Uo{\alpha}\right)^t \Ue{\alpha}\vec{a}.
\end{eqnarray}
Note the difference between the definitions for even and odd times, which is a
consequence of the staggering of time evolution. Time-evolution of local observables
is particularly simple as one only has to consider the subsection of the chain on which
the observables act nontrivially. This follows from the fact that the time-evolution
operator preserves the identity observable, which implies the following,
\begin{eqnarray}
  U^{\alpha}_{x-1,x,x+1} \oo{q_1 q_2\ldots q_r}_j = \oo{q_1 q_2 \ldots q_r}_j,
  \qquad \text{for } x<j \text{ or } x>j+r-1.
\end{eqnarray}

\section{Ergodicity and local conserved quantities}
Besides the seemingly small differences in the collision rules for simple
particles, the two models exhibit qualitatively very different dynamical
features: the automaton given by~$\chi_1$ behaves as an integrable system,
while~$\chi_2$ appears to be non-integrable. In this section we provide evidence
supporting this characterization.

\begin{figure}
  \centering
  \begin{tikzpicture}[baseline={([yshift=-0.5ex]current bounding box.south)}]
    \draw[thick,border,->] ({-\a},{-\a}) -- ({-\a},{16*\a}) node[midway,left]
    {$t$};
    \node at ({-\a},{8.5*\a}) {$\phantom{t}$};
  \end{tikzpicture}%
  \ \includegraphics[width=0.19\textwidth]{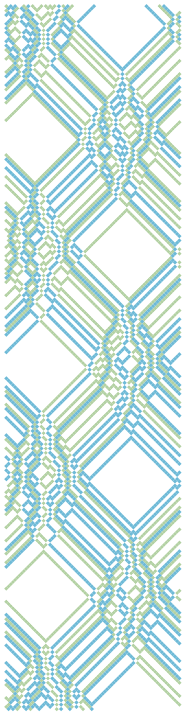}%
  \qquad \includegraphics[width=0.19\textwidth]{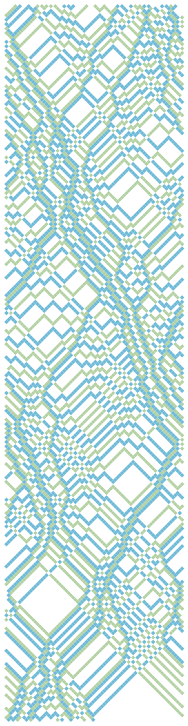}\\
  \includegraphics[width=0.66\textwidth]{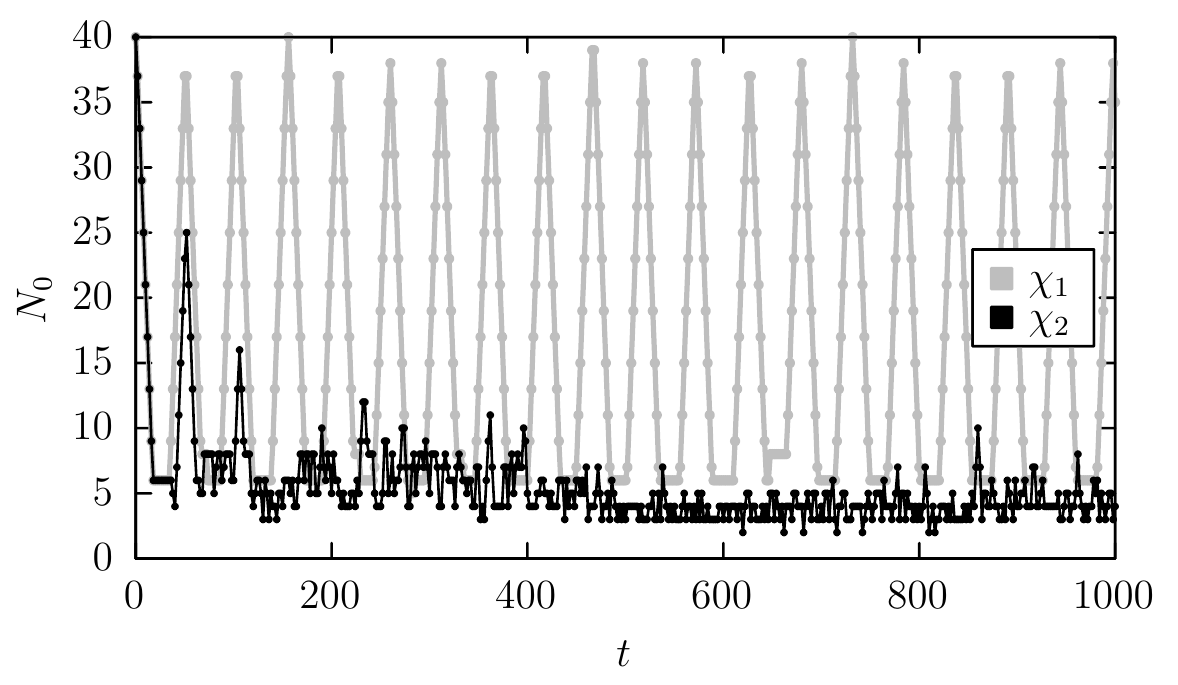}
  \caption{\label{fig:ergodicity}
  Time evolution of a half-empty configuration with periodic boundary
  conditions. The initial configuration consists of the left half initialized
  in a randomly selected configuration of 40 sites, and the right half that is
  initially empty. The system is then left to evolve according to
  $\chi_1$ and $\chi_2$ in the left and right top image respectively. The plot
  at the bottom shows the time dependence of the maximal number of consecutive $0$
  in a configuration, $N_0$. This provides a quantitative description of the 
  behaviour seen from the images of trajectories on larger time scales. In the
  case of~$\chi_1$ the quantity $N_0$ shows periodic behaviour that persists on
  time scales shown here, while for~$\chi_2$ the oscillations disappear very quickly.
  The initial configuration is the same in all three figures.
  }
\end{figure}
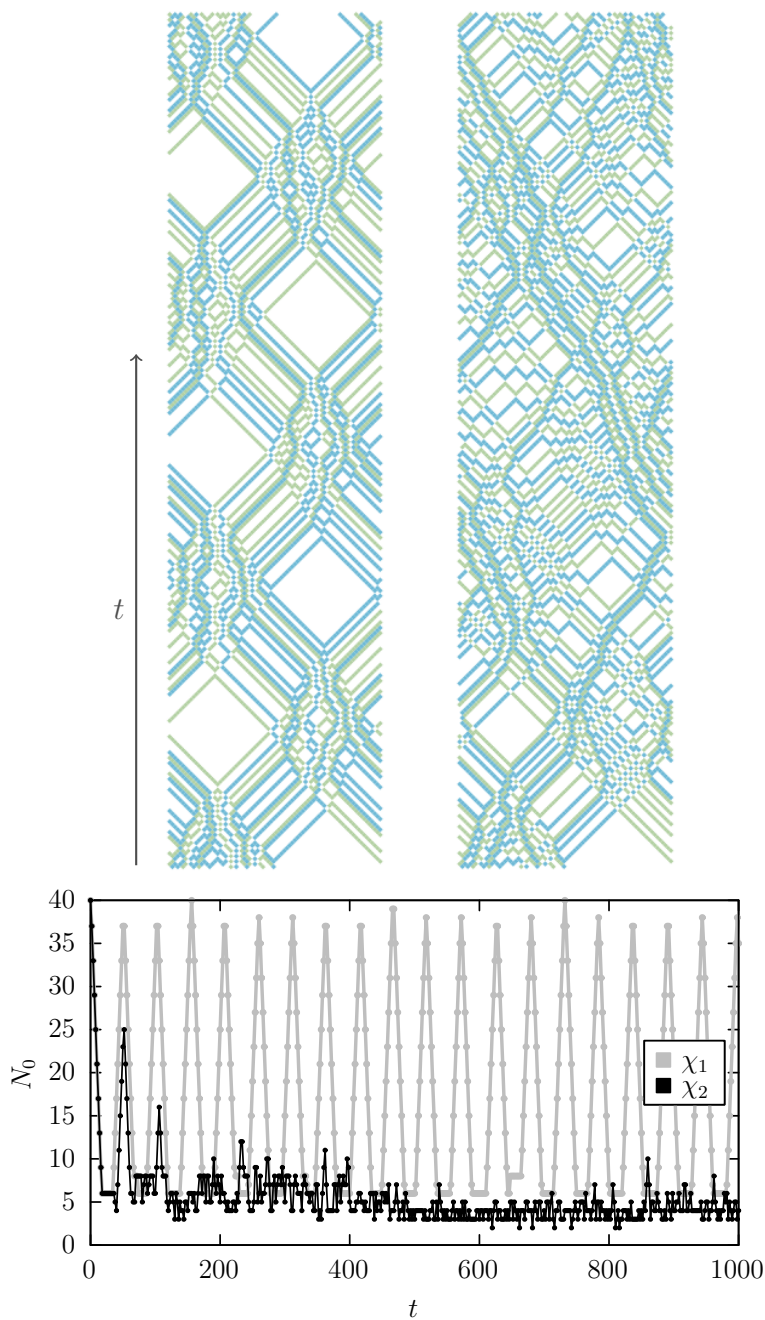

Let us start by considering the dynamics of a simple initial state,
where half of the chain is initialized in a randomly chosen configuration
and the second half is empty. Evolving this configuration in time, with the assumption
of periodic boundaries, the trajectory in the case of~$\chi_1$ is very regular, with almost
periodically appearing gaps, while in the trajectory of~$\chi_2$ this behaviour 
very quickly disappears, as is shown in Figure~\ref{fig:ergodicity}. The dynamics
of~$\chi_1$ therefore appears to be nonergodic, and suggests the existence of
additional local conservation laws.

Local conserved quantities are observables~$Q$ preserved under time evolution,
\begin{eqnarray}
  \Uo{\alpha} \Ue{\alpha} Q = Q,
\end{eqnarray}
that can be expressed as a linear combination of strictly local terms $q$ with
finite maximum support~$r>0$,
\begin{eqnarray}
  Q=\sum_{j=1}^{n} \eta_{2j}(q),\qquad
  \eta_{x}(\oo{s_1 s_2 \ldots s_k}_j)=
  \oo{s_1 s_2 \ldots s_k}_{j+x}.
  \label{eq:defQ}
\end{eqnarray}
Note that due to the staggering of time-evolution we assume invariance under
translations for \emph{even} numbers of sites. The three most local conserved
quantities can be found easily and are the same for both models,
\begin{eqnarray}\label{eq:simpleCQs}
  \fl
  Q_1=\sum_{j=1}^{2n} (-1)^{j}\left(\oo{00}_j + \oo{11}_j+\oo{22}_j\right),\qquad
  Q_2=\sum_{j=1}^{2n} \oo{12}_j,\qquad
  Q_3=\sum_{j=1}^{2n} \oo{21}_j.
\end{eqnarray}
Despite the fact that the charges involve only nearest neighbour couplings, their densities are supported on three adjacent sites because of the staggering (\ref{eq:defQ})
\begin{equation}
q_1 = \sum_{p,s=0}^2
([pps]-[spp]),\quad
q_2 = \sum_{s=0}^2 ([12s]+[s 12]),\quad
q_3 = \sum_{s=0}^2 ([21s]+[s 21]).
\end{equation}
The quantities~$Q_2$ and $Q_3$ represent the total number of pairs of consecutive
sites occupied by pairs~$1$, $2$ and $2$, $1$ respectively, while $Q_1$ is measuring
the difference between numbers of simple (unit width) solitons that move to the right and to
the left.

However, these are not necessarily all of the local conservation laws. For larger 
(but finite) lengths of the support~$r$, we can numerically search for conserved
observables by diagonalizing the time-evolution operator reduced to the space of 
extensive local observables, as described in Ref.~\cite{prosen2007chaos}. The numerical
results reported in Table~\ref{tab:numCQs}, suggest that~\eqref{eq:simpleCQs} is 
the complete set of the local conservation laws for~$\chi_2$.  However, in the
case of~$\chi_1$ we observe (at least for accessible support lengths~$r\le 8$) that
the number of conserved quantities increases linearly with the support. In
particular, the system exhibits exactly one conserved quantity with exact
support~$r>3$ (see~\ref{app:consQuant} for the explicit form of the quantity
with support~$r=4$). Note that the linear increase of the number of conservation
laws with the support is typical of Bethe-Ansatz/Yang-Baxter integrable spin chains~\cite{faddeev1996algebraic}
and is very different to the situation observed in RCA54, where the number of
conservation laws grows \emph{exponentially} with the system size~\cite{klobas2020exact}.
\begin{table}
  \centering
  \begin{tabular}{r || c | c | c | c | c | c }
    $r$       & 3 & 4 & 5 & 6 & 7 & 8 \\ \hhline{= =:=:=:=:=:=}
    $\#_1(r)$ & 3 & 4 & 5 & 6 & 7 & 8 \\ \hline
    $\#_2(r)$ & 3 & 3 & 3 & 3 & 3 & 3 
  \end{tabular}
  \caption{\label{tab:numCQs} Numbers of local conserved quantities. The reported values $\#_1(r)$ and~$\#_2(r)$ are numbers of linearly independent local extensive
  observables with the support at most $r$ that are invariant under the time
  evolution given by $\chi_1$ and $\chi_2$, respectively.
  }
\end{table}

The automaton given by the dynamical rule~$\chi_1$ therefore appears to be
integrable, while~$\chi_2$ is not. Even though we cannot provide a proof for
this, we will occasionally discriminate between the two models by referring to
them as (non)integrable.  

\section{Matrix product state representation of Gibbs states}\label{sec:GibbsStates}
Analogously to RCA54 and the deterministic PXP
model (rule 201)~\cite{prosen2017exact,wilkinson2020exact}, the two-species automata exhibit
a family of stationary states that can be efficiently expressed in terms of
products of matrices satisfying a cubic algebraic relation.

A stationary state~$\vec{p}_{\alpha}$ of the automaton determined
by~$\chi_{\alpha}$ should be invariant under evolution for two time-steps,
\begin{eqnarray}\label{eq:stationaryCondition1}
  \vec{p}_{\alpha} = \Uo{\alpha}\Ue{\alpha}\vec{p}_{\alpha}.
\end{eqnarray}
We introduce two versions of the state, $\vec{p}_{\alpha}$ and
$\vec{p}_{\alpha}^{\prime}$, that are mapped into each other under the even and
the odd time-evolution operator,
\begin{eqnarray}\label{eq:stationaryCondition}
  \vec{p}_{\alpha}^{\prime}=\Ue{\alpha}\vec{p}_{\alpha},\qquad
  \vec{p}_{\alpha}=\Uo{\alpha}\vec{p}_{\alpha}^{\prime}.
\end{eqnarray}

A class of states that satisfies this condition can be expressed as a
\emph{matrix product state} (MPS) by defining~$\vec{W}_j(\xi,\omega,\lambda)$,
$\vec{W}^{\prime}_j(\xi,\omega,\lambda)$ as vectors in the physical space
associated to the site~$1\le j \le 2n$,
\begin{eqnarray}
  \vW_j(\xi,\omega,\lambda) =
  \begin{bmatrix}
    \W_0(\xi,\omega,\lambda)\\
    \W_1(\xi,\omega,\lambda)\\
    \W_2(\xi,\omega,\lambda)\\
  \end{bmatrix}_j,\qquad
  \vV_j(\xi,\omega,\lambda) =
  \begin{bmatrix}
    \V_0(\xi,\omega,\lambda)\\
    \V_1(\xi,\omega,\lambda)\\
    \V_2(\xi,\omega,\lambda)\\
  \end{bmatrix}_j,
\end{eqnarray}
while their components are the following $3\times 3$ matrices,
\begin{eqnarray}\fl \label{eq:GibbsRepresentation}
  \eqalign{
    \W_0(\xi,\omega,\lambda)=\begin{bmatrix}
    \xi&1&1\\
    0&0&0\\
    0&0&0
  \end{bmatrix},\quad
  \W_1(\xi,\omega,\lambda)=\begin{bmatrix}
    0&0&0\\
    1&\xi&\omega\\
    0&0&0
  \end{bmatrix},\quad
  \W_2(\xi,\omega,\lambda)=\begin{bmatrix}
    0&0&0\\
    0&0&0\\
    1&\lambda&\xi
  \end{bmatrix},\\
  \V_s(\xi,\omega,\lambda)=\W_s(\frac{1}{\xi},\omega,\lambda),\quad s\in\{0,1,2\}.
}
\end{eqnarray}

These matrices together with the update rules~$\chi_1$, $\chi_2$ satisfy the
following cubic algebraic relations,~\footnote{For compactness of notation, we
will often suppress the explicit dependence to parameters~$\xi,\omega,\lambda$,
i.e.\ $W^{(\prime)}_s (\xi,\omega,\lambda) \to W^{(\prime)}_s$.}
\begin{eqnarray}
  \fl
  \eqalign{
  W_{s_1} \V_{\chi_1(s_1,s_2,s_3)} \W_{s_3} = \V_{s_1} \W_{s_2} \W_{s_3},\qquad
  W_{s_1} \V_{\chi_2(s_1,s_2,s_3)} \W_{s_3} = \V_{s_1} \W_{s_2} \W_{s_3},\\
  W_{s_1} \V_{\chi_1(s_1,s_2,s_3)} \V_{s_3} = \V_{s_1} \W_{s_2} \V_{s_3},\qquad
  W_{s_1} \V_{\chi_2(s_1,s_2,s_3)} \V_{s_3} = \V_{s_1} \W_{s_2} \V_{s_3},
}
\end{eqnarray}
for any combination of~$s_1, s_2, s_3\in\{0,1,2\}$. 
This can be compactly summarized in the vectorized 
form as
\begin{eqnarray} \label{eq:algebraicSimple}
  U^{\alpha}_{123} \vW_1 \vV_2 \vW_3 = \vV_1 \vW_2 \vW_3,\qquad
  U^{\alpha}_{123} \vW_1 \vV_2 \vV_3 = \vV_1 \vW_2 \vV_3,
\end{eqnarray}
where~$\alpha\in\{1,2\}$ discriminates between the two models. Note that
the second relation follows from the first one from 
$U^{\alpha} = (U^{\alpha})^{-1}$ and the mapping between the
primed and unprimed matrices,
$\W_{s} \xleftrightarrow{\xi\leftrightarrow \xi^{-1}}\V_{s}$.

The immediate consequence of the algebraic relation is that the following
family of states is stationary in \emph{both} models, 
\begin{eqnarray}\fl
  \vec{p}(\xi,\omega,\lambda)=\frac{1}{Z}
  \tr\left(\vW_1 \vV_2 \vW_3 \cdots \vV_{2n}\right),\quad
  \vec{p}^{\prime}(\xi,\omega,\lambda)=\frac{1}{Z}
  \tr\left(\vV_1 \vW_2 \vV_3 \cdots \vW_{2n}\right),
\end{eqnarray}
since for both~$\alpha=1$ and~$\alpha=2$, the states~$\vec{p}$ and~$\vec{p}^{\prime}$
satisfy the condition~\eqref{eq:stationaryCondition}. The prefactor~$Z$ is determined
by the normalization condition,
\begin{eqnarray}
  Z=\tr\Big((\W_0+\W_1+\W_2)(\V_0+\V_1+\V_2)\Big)^n,
\end{eqnarray}
and the length of the chain is assumed to be~$2n$.

The set of vectors~$\vec{p}(\xi,\omega,\lambda)$ can be interpreted as a family of
\emph{Gibbs states}. To see that, we first note that the structure of the 
matrices~$\W_s,\V_s$ implies the following,
\begin{eqnarray}
  \fl
  \eqalign{
  \W_{s} \hre_{b} = \hre_{s}
  \left(\delta_{s,b}(\xi-1) + \delta_{s,1}\delta_{b,2}(\omega-1)
  + \delta_{s,2}\delta_{b,1}(\lambda-1)+1\right),\\
  \V_{s} \hre_{b} = \hre_{s}
  \left(\delta_{s,b}(\xi^{-1}-1) + \delta_{s,1}\delta_{b,2}(\omega-1)
  + \delta_{s,2}\delta_{b,1}(\lambda-1)+1\right),}
  \qquad
  \hre_s=
  \begin{bmatrix}
    \delta_{s,0}\\ \delta_{s,1}\\ \delta_{s,2}
  \end{bmatrix},
\end{eqnarray}
where~$\{\hre_s\}_{s=0,1,2}$ is the canonical basis of the auxiliary space.
Comparing the coefficients and components in these relations with the explicit
form of conserved quantities~\eqref{eq:simpleCQs}, one immediately realizes that the
components of the probability distribution~$\vec{p}(\xi,\omega,\lambda)$
corresponds to the probabilities of configurations in the Gibbs ensemble,
\begin{eqnarray}
  p_{\underline{s}}=\frac{1}{Z} \tr\left(\W_{s_1}\V_{s_2}\W_{s_3}\cdots\V_{s_{2n}}\right)
  \propto
  \re^{\mu_1 Q_1(\underline{s})+\mu_2 Q_2(\underline{s})+\mu_3 Q_3(\underline{s})},
  %\xi^{Q_1(\underline{s})} \omega^{Q_2(\underline{s})} \lambda^{Q_3(\underline{s})}.
\end{eqnarray}
where we identify chemical potentials~$\mu_1,\mu_2,\mu_3$ as
\begin{eqnarray}
  \mu_1=\log\xi,\qquad
  \mu_2=\log\omega,\qquad
  \mu_3=\log\lambda.
\end{eqnarray}
This explains why the same state~$\vec{p}$ is stationary in both models: the
first three conserved quantities are common to both automata and therefore they
exhibit the same simple Gibbs state. Note that in the case of~$\chi_2$, we
expect this to be the complete family of equilibrium states, while for~$\chi_1$ one
should include the (presumably) infinite set of local charges to obtain the most general family of generalised Gibbs states (GGE).

\section{Stationary states with higher conservation laws} \label{sec:statesHigherCQs}
Since the integrable model exhibits more than $3$ conservation laws, it is possible to find
richer GGEs that take into account also some of the conserved quantities with larger support.
Here we show an example with $4$ chemical potentials: in addition to
$\log \xi$, $\log \omega$ and $\log \lambda$ we introduce also $\log \mu$, which corresponds
to the conserved charge with support $4$. Throughout the section we mostly consider
the automaton given by the set of rules $\chi_1$, and at the end we also discuss some related observations for $\chi_2$.

\subsection{Patch state ansatz} \label{sec:PSA}
The first form of the stationary state we consider is the \emph{patch-state ansatz}, as
introduced in~\cite{prosen2016integrability}. We assume that the components of the
translationally invariant steady state $\vec{p}$ can be written in the following form,
\be
p_{(s_1,s_2,\ldots,s_{2n})} =\frac{1}{Z}
t_{s_1 s_2 s_3 s_4} t_{s_3 s_4 s_5 s_6}\cdots
t_{s_{2n-3}s_{2n-2}s_{2n-1}s_{2n}} t_{s_{2n-1}s_{2n}s_{1}s_{2}},
\ee
where $Z$ is a normalization constant and $t$ is a tensor with components $t_{s_1 s_2 s_3 s_4}$
that are determined so that the fixed-point condition~\eqref{eq:stationaryCondition1} is
satisfied. In particular, the form of the stationarity condition that is most convenient for
us is 
\be
\Ue{1} \vec{p} =\Uo{1} \vec{p},
\ee
which is in terms of the components rewritten as,
\be
p_{(s_1 \chi(s_1,s_2,s_3) s_3,\chi(s_3,s_4,s_5)\ldots\chi(s_{2n-1},s_{2n},s_1))}=
p_{(\chi(s_{2n},s_1,s_2)s_2\chi(s_2,s_3,s_4)s_5\ldots s_{2n})},
\ee
for each configuration $\ul{s}=(s_1,s_2,\ldots,s_{2n})\in\mathbb{Z}_{3}^{2n}$.
Up to a gauge transformation
\be
    t_{s_1 s_2 s_3 s_4} \to f_{s_1 s_2} t_{s_1 s_2 s_3 s_4} f^{-1}_{s_3 s_4},
\ee
we find a unique~\footnote{We assume all of the entries to be nonvanishing,
which means that each configuration $\ul{s}$ arises with a non-zero probability
$p_{\ul{s}}>0$.}
four-parameter family of solutions (to be proven below) that can be 
compactly summarized in terms of the $9\times 9$ transfer matrix $T$ with components,
$T_{(s_1 s_2),(s_1^{\prime},s_2^{\prime})}=t_{s_1 s_2 s_1^{\prime} s_2^{\prime}}$,
\be
T(\xi,\omega,\lambda,\mu) = 
\begin{bmatrix}
    1 & 1 & 1 & 1 & 1 & 1 & 1 & 1 & 1 \\
    \mu  & \mu  & 1 & \frac{1}{\xi^2}  & \frac{\mu}{\xi^2}  & \frac{1}{\xi^2}  
    & \frac{\omega}{\xi}  & \frac{\omega}{\xi} & \frac{\mu\omega}{\xi}\\
    \mu  & 1 & \mu  & \frac{\lambda}{\xi} & \frac{\lambda\mu}{\xi} & \frac{\lambda}{\xi} &
    \frac{1}{\xi^2} & \frac{1}{\xi^2} & \frac{\mu}{\xi^2}\\
    \mu  & \mu  & 1 & \mu  & \mu  & \mu  & 1 & 1 & \mu \\
    \xi^2 & \xi^2 & \xi^2 & 1 & \mu  & 1 & \xi\omega & \xi\omega & \xi\omega \\
    \mu\xi\omega & \xi\omega & \mu\xi\omega & \lambda\omega & \lambda\mu\omega  &
    \lambda\omega  & \frac{\omega}{\xi} & \frac{\omega}{\xi} & \frac{\mu\omega}{\xi} \\
    \mu  & 1 & \mu  & 1 & \mu  & 1 & \mu  & \mu  & \mu  \\
    \lambda\mu\xi & \lambda\mu\xi & \lambda\xi & \frac{\lambda}{\xi} & \frac{\lambda\mu}{\xi} &
    \frac{\lambda}{\xi} & \lambda\omega  & \lambda\omega & \lambda\mu\omega\\
    \xi^2 & \xi^2 & \xi^2 & \lambda\xi & \lambda\xi & \lambda\xi & 1 & 1 & \mu
\end{bmatrix}.
\ee
As before, we can define two versions of the state, $\vec{p}$ and $\vec{p}^{\prime}$, that
get mapped into each other under the evolution for one time-step
(cf.~\eqref{eq:stationaryCondition}). In our case the odd-time-step version of the stationary
state $\vec{p}^{\prime}$ is $\vec{p}$ shifted for one site with the components of the primed
state explicitly given by
\be
p^{\prime}_{(s_1 s_2 \ldots s_{2n})}
=p_{(s_2 \ldots s_{2n} s_1)} = 
t_{s_2 s_3 s_4 s_5} t_{s_4 s_5 s_6 s_7}\cdots t_{s_{2n} s_1 s_2 s_3}.
\ee
To show that~\eqref{eq:stationaryCondition} really holds for $\vec{p}$ and $\vec{p}^{\prime}$
defined above, we only need to verify that the tensor $t$ satisfies a finite set of local
relations. We start by explicitly rewriting the condition for a finite length of the chain,
say $2n=6$,
\be 
\eqalign{
    t_{s_1 s_2 s_3 s_4} t_{s_3 s_4 s_5 s_6} t_{s_5 s_6 s_1 s_2}\\
    =\mkern-5mu
    t_{\chi_1(s_1,s_2,s_3)s_3\chi_1(s_3,s_4,s_5)s_5} 
    t_{\chi_1(s_3,s_4,s_5)s_5\chi_1(s_5,s_6,s_1)s_1}
    t_{\chi_1(s_5,s_6,s_1)s_1\chi_1(s_1,s_2,s_3)s_3}\\
    =\mkern-5mu
    t_{s_2\chi_1(s_2,s_3,s_4)s_4\chi_1(s_4,s_5,s_6)} 
    t_{s_4\chi_1(s_4,s_5,s_6)s_6\chi_1(s_6,s_1,s_2)} 
    t_{s_6\chi_1(s_6,s_1,s_2)s_2\chi_1(s_2,s_3,s_4)}.
}
\ee
The proof of~\eqref{eq:stationaryCondition} for the chain of length $2n=6$ amounts
to checking that the above relations are fulfilled for all $3^6$ configurations. To
extend the proof to an arbitrary length, we observe that the following set of compatibility
conditions is also fulfilled,
\be
\eqalign{
\frac{t_{s_1 s_2 s_3 s_4}t_{s_3 s_4 s_5 s_6}}{t_{s_1 s_2 s_5 s_6}}\\
=\mkern-5mu\frac{t_{s_0 \chi_1(s_0,s_1,s_2) s_2 \chi_1(s_2,s_3,s_4)}
       t_{s_2 \chi_1(s_2,s_3,s_4) s_4 \chi_1(s_4,s_5,s_6)} 
       t_{s_4 \chi_1(s_4,s_5,s_6) s_6 \chi_1(s_6,s_7,s_8)}}
      {t_{s_0 \chi_1(s_0,s_1,s_2) s_2 \chi_1(s_2,s_5,s_6)}
       t_{s_2 \chi_1(s_2,s_5,s_6) s_6 \chi_1(s_6,s_7,s_8)}}\\
=\mkern-5mu\frac{t_{\chi_1(s_{-1},s_0,s_1) s_1 \chi_1(s_1,s_2,s_3) s_3}
       t_{\chi_1(s_1,s_2,s_3)    s_3 \chi_1(s_3,s_4,s_5) s_5}
       t_{\chi_1(s_3,s_4,s_5)    s_5 \chi_1(s_5,s_6,s_7) s_7}}
      {t_{\chi_1(s_{-1},s_0,s_1) s_1 \chi_1(s_1,s_2,s_5) s_5}
       t_{\chi_1(s_1,s_2,s_5)    s_5 \chi_1(s_5,s_6,s_7) s_7}}.
   }
\ee
Even though these two sets of relations are rather large and checking their
validity cannot be easily performed by hand (as they involve $2\cdot 3^9+2\cdot
3^6=40824$ separate equations), the number of relevant relations is still
finite and they can be easily verified using computer algebra systems. From
here, Eq.~\eqref{eq:stationaryCondition} immediately follows and the pair
of states $(\vec{p},\vec{p}^{\prime})$ is therefore invariant under time-evolution.

\subsection{Equivalent MPS}
Another, perhaps more convenient, form of the stationary state~$\vec{p}$ introduced above
is an MPS, defined analogously to the Gibbs state from Section~\ref{sec:GibbsStates},
\be \label{eq:defWVlarger}
\vec{p}=\frac{1}{Z}
\tr\left(\vW_1 \vV_2 \cdots \vV_{2n}\right),\qquad
\vec{p}=\frac{1}{Z}
\tr\left(\vW_1 \vV_2 \cdots \vV_{2n}\right).
\ee
However, since the stationary state is now richer, the matrices $\W_s$, $\V_s$ now have
to be larger. In particular, we find a $7\times 7$ dimensional representation, 
for which the state~\eqref{eq:defWVlarger} exactly corresponds to the patch state ansatz from the previous
subsection,
\be \fl
\eqalign{
\W_0=
\begin{bmatrix}
    1 & 0 & 0 & 0 & 0 & 0 & 0 \\
    0 & 1 & 0 & 0 & 0 & 0 & 0 \\
    0 & 0 & 1 & 0 & 0 & 0 & 0 \\
    0 & 0 & 0 & 0 & 0 & 0 & 0 \\
    0 & 0 & 0 & 0 & 0 & 0 & 0 \\
    0 & 0 & 0 & 0 & 0 & 0 & 0 \\
    0 & 0 & 0 & 0 & 0 & 0 & 0 \\
\end{bmatrix},
&\V_0=
\begin{bmatrix}
    1 & 1 & 1 & 1 & 1 & 1 & 1 \\
    0 & 0 & 0 & 0 & 0 & 0 & 0 \\
    0 & 0 & 0 & 0 & 0 & 0 & 0 \\
    \mu  & \mu  & 1 & \mu  & \mu  & 1 & \mu  \\
    0 & 0 & 0 & 0 & 0 & 0 & 0 \\
    \mu  & 1 & \mu  & 1 & \mu  & \mu  & \mu  \\
    0 & 0 & 0 & 0 & 0 & 0 & 0 \\
\end{bmatrix},
\\
\W_1 =
\begin{bmatrix}
    0 & 0 & 0 & 0 & 0 & 0 & 0 \\
    0 & 0 & 0 & 0 & 0 & 0 & 0 \\
    0 & 0 & 0 & 0 & 0 & 0 & 0 \\
    0 & 0 & \xi\omega & 1 & 0 & 0 & 0 \\
    0 & 0 & 0 & 0 & 1 & 0 & 0 \\
    0 & 0 & 0 & 0 & 0 & 0 & 0 \\
    0 & 0 & 0 & 0 & 0 & 0 & 0
\end{bmatrix}, \qquad
&\V_1 = 
\begin{bmatrix}
    0 & 0 & 0 & 0 & 0 & 0 & 0 \\
    \mu&\mu&1&\frac{1}{\xi^2}&\frac{\mu}{\xi^2}&\frac{\omega}{\xi}&\frac{\mu\omega}{\xi}\\
    0 & 0 & 0 & 0 & 0 & 0 & 0 \\
    0 & 0 & 0 & 0 & 0 & 0 & 0 \\
    \xi^2&\xi^2&\xi^2&1 & \mu  & \xi\omega & \xi\omega \\
    0 & 0 & 0 & 0 & 0 & 0 & 0 \\
    0 & 0 & 0 & 0 & 0 & 0 & 0 \\
\end{bmatrix},\\
\W_2 =
\begin{bmatrix}
    0 & 0 & 0 & 0 & 0 & 0 & 0 \\
    0 & 0 & 0 & 0 & 0 & 0 & 0 \\
    0 & 0 & 0 & 0 & 0 & 0 & 0 \\
    0 & 0 & 0 & 0 & 0 & 0 & 0 \\
    0 & 0 & 0 & 0 & 0 & 0 & 0 \\
    0 & \lambda\xi & 0 & 0 & 0 & 1 & 0 \\
    0 & 0 & 0 & 0 & 0 & 0 & 1 
\end{bmatrix},
&\V_2 = 
\begin{bmatrix}
    0 & 0 & 0 & 0 & 0 & 0 & 0 \\
    0 & 0 & 0 & 0 & 0 & 0 & 0 \\
    \mu  & 1 & \mu  & \frac{\lambda}{\xi} & \frac{\lambda\mu}{\xi} &
    \frac{1}{\xi^2}  & \frac{\mu}{\xi^2} \\
    0 & 0 & 0 & 0 & 0 & 0 & 0 \\
    0 & 0 & 0 & 0 & 0 & 0 & 0 \\
    0 & 0 & 0 & 0 & 0 & 0 & 0 \\
    \xi^2 & \xi^2 & \xi^2 & \lambda\xi & \lambda\xi & 1 & \mu
\end{bmatrix}.
}
\ee

To independently verify the stationarity of~\eqref{eq:defWVlarger}, we identify
a set of cubic algebraic relations that generalises
Eq.~\eqref{eq:algebraicSimple},
\be \label{eq:WWW}
U^{1}_{123} \vX_1 \vW_2 \vV_3 = \vW_1\vV_2\vX_3,\qquad
U^{1}_{123} \vY_1 \vV_2 \vW_3 = \vV_1\vW_2\vY_3,
\ee
where the components of $\vX$, $\vY$ are $7\times 7$ matrices given in~\ref{sec:matXY}.
To prove that the MPS built out of these matrices is stationary for periodic boundaries,
one needs to also take into account the following set of easily verifiable compatibility
relations,
\be \label{eq:WWWcompt}
U^{1}_{123}\vX_{1} \vW_2 \vY_3 = \vW_2 \vV_2 \vW_3,\qquad
U^{1}_{123}\vY_{1} \vV_2 \vX_3 = \vV_1 \vW_2 \vV_3.
\ee

\subsection{The second model}
For the `non-integrable' rule $\chi_2$, one needs to fix $\mu=1$ and the state remains
stationary. However, the precise form of local algebraic relations changes slightly. 
The identities on the right of~\eqref{eq:WWW} and~\eqref{eq:WWWcompt} remain valid, 
\be \label{eq:WWW2} \fl
\eqalign{
U^2_{123}\left.\vY_1 \vV_2 \vW_3\right|_{\mu\to 1}\mkern-4mu=\mkern-4mu
\left.\vV_1 \vW_2 \vY_3\right|_{\mu\to 1},\qquad
U^2_{123}\left.\vY_1 \vV_2 \vX_3\right|_{\mu\to 1}\mkern-4mu=\mkern-4mu
\left.\vV_1 \vW_2 \vV_3\right|_{\mu\to 1},}
\ee
while the other two identities do not hold directly, but have to be multiplied
by $\V_{s}$ from the left for any $s\in\{0,1,2\}$, so that the relation becomes
effectively quartic rather than cubic,
\be
\eqalign{
U^{2}_{234}\left.\vV_{1} \vX_2 \vW_3 \vV_4\right|_{\mu\to 1}=
\left.\vV_1 \vW_2 \vV_3 \vX_4\right|_{\mu\to 1},\\
U^2_{234}
\left.\vV_1 \vX_2 \vW_3 \vY_4\right|_{\mu\to 1}=
\left.\vV_1 \vW_2 \vV_3 \vW_4\right|_{\mu\to 1}.}
\label{eq:WWWcompt2}
\ee
Using the modified set of identities, one can prove that $\vV$ and $\vW$ in the
case of $\mu=1$ provide a stationary state also for the non-integrable model.
One might wonder if there is another $7\times 7$ representation, for which the
limit $\mu\to 1$ gives directly the stationary state also for $\chi_2$ without
resorting to the modified algebra.  At the moment such a representation has
still not been found, and it is not clear whether it exists or not.

Nonetheless, since we expect that the case $\mu=1$ corresponds to the Gibbs state,
the matrices~$\W_s$, $\V_s$ should in this limit reduce to the $3\times 3$ representation
introduced in Section~\ref{sec:GibbsStates}, which obeys a simple cubic algebra for
\emph{both} models. To prove this, we introduce the following set of $3\times
7$ and $7\times 3$ matrices,
\be  \fl
\eqalign{
S_1=
\begin{bmatrix}
    1&0&0\\
    0&\frac{1}{\xi}&0\\
    0&0&\frac{1}{\xi}\\
    1&0&0\\
    0&\xi&0\\
    1&0&0\\
    0&0&\xi
\end{bmatrix},
&R_2=
\begin{bmatrix}
    \frac{1}{2\xi} & 0 & 0 \\
    \frac{1}{4\xi} & 0 & 0 \\
    \frac{1}{4\xi} & 0 & 0 \\
    0 & \frac{1}{2} & 0 \\
    0 & \frac{1}{2} & 0 \\
    0 & 0 & \frac{1}{2} \\
    0 & 0 & \frac{1}{2}
\end{bmatrix},\\
S_2=
\begin{bmatrix}
    \frac{1}{4}&0&0&\frac{1}{4}&0&\frac{1}{2}&0\\
    0&\frac{\xi}{1-\xi^2}&0&0&\frac{\xi}{\xi^2-1}&0&0\\
    0&0&\frac{\xi}{1-\xi^2}&0&0&0&\frac{\xi}{\xi^2-1}
\end{bmatrix},
\qquad
&R_1=
\begin{bmatrix}
    \xi&\xi&\xi&0&0&0&0\\
    0&0&0&1&1&0&0\\
    0&0&0&0&0&1&1
\end{bmatrix},
}
\ee
which for $\mu=1$ satisfy the following set of relations for any pair
of $s,s^{\prime}\in\{0,1,2\}$,
\be
\W_{s}S_1 S_2 \V_{s^{\prime}}=\W_s \V_{s^{\prime}},\qquad
\V_{s}R_2 R_1 \W_{s^{\prime}}=\V_s \W_{s^{\prime}}.
\ee
This immediately implies that by the following transformation
\be
\left.R_1 \W_s S_1\right|_{\mu\to 1} \mapsto \W_s
,\qquad
\left. S_2 \V_s R_2 \right|_{\mu\to 1}\mapsto \V_s,
\ee
we obtain precisely the set of $3\times 3$ matrices~\eqref{eq:GibbsRepresentation}.

\section{Spatio-temporal correlation functions}
Let us now consider dynamical correlation functions of local densities of conserved
quantities. In particular, we define~$C(x,t)$ as the following correlation function,
\begin{eqnarray}\label{eq:defCxt}
  C(x,t)=\expval*{\left.q^{(+)}_x\right.^t q^{(+)}_0}_{\vec{p}}-
  \expval*{q_0^{(+)}}_{\vec{p}}^2,
  \qquad
  q^{(+)}_j=\oo{12}_j+\oo{21}_j,
\end{eqnarray}
where the state~$\vec{p}$ is assumed to be stationary (i.e.\ invariant under
time evolution).
For simplicity this is the only correlation
function we discuss here, however the results do not strongly depend on the exact choice
of the correlation function in question. Note that $q^{(+)}$ is the local density of the
sum of~$Q_2$ and~$Q_3$,
\begin{eqnarray}
  \sum_{j=1}^{2n} q_{j}^{(+)}=Q_2+Q_3.
\end{eqnarray}
In this section, the position label is bounded by $\pm t$ (and not
$1$ and $2n$ as elsewhere), and the full system size at time~$t$ is assumed to be
much larger than~$2t$. In this limit the expectation values of local
observables in the Gibbs states can be expressed in terms of MPS defined on a
finite length of the chain, as described in~\ref{app:gibbsHDL}.

\begin{figure}
  \centering
  \includegraphics[width=\textwidth]{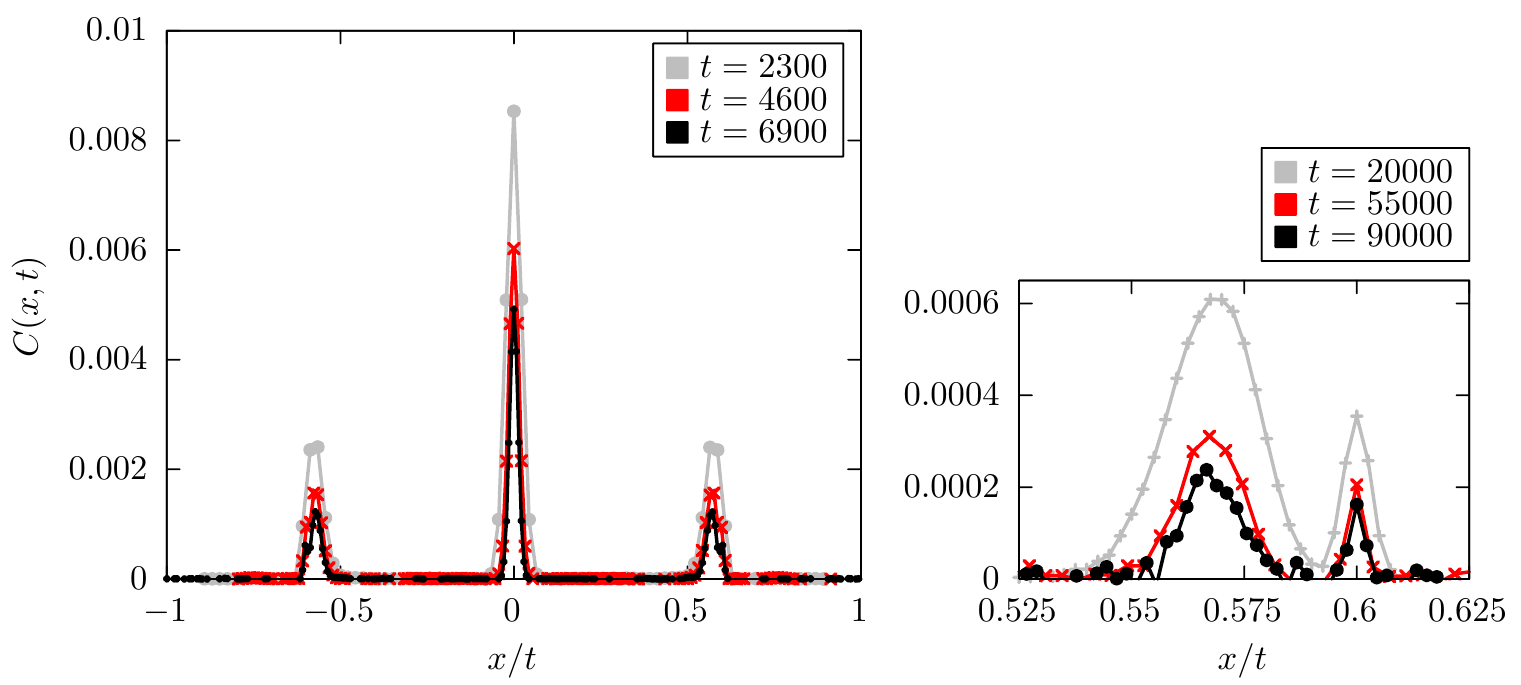}
  \caption{\label{fig:IntegrableCorrelations} Dynamical correlation function~$C(x,t)$
  for the \emph{integrable} case and $\vec{p}=\vec{p}_{\infty}$. At short times
  the correlation function appears to exhibit three peaks: a central one and
  two side peaks moving to the left and right. However, at longer times it
  becomes evident that each of the side peaks separates into two distinct peaks
  that move with very close but different velocities, as can be seen in the closeup
  on the right.
  }
\end{figure}

\begin{figure}
  \centering
  \includegraphics[width=0.8\textwidth]{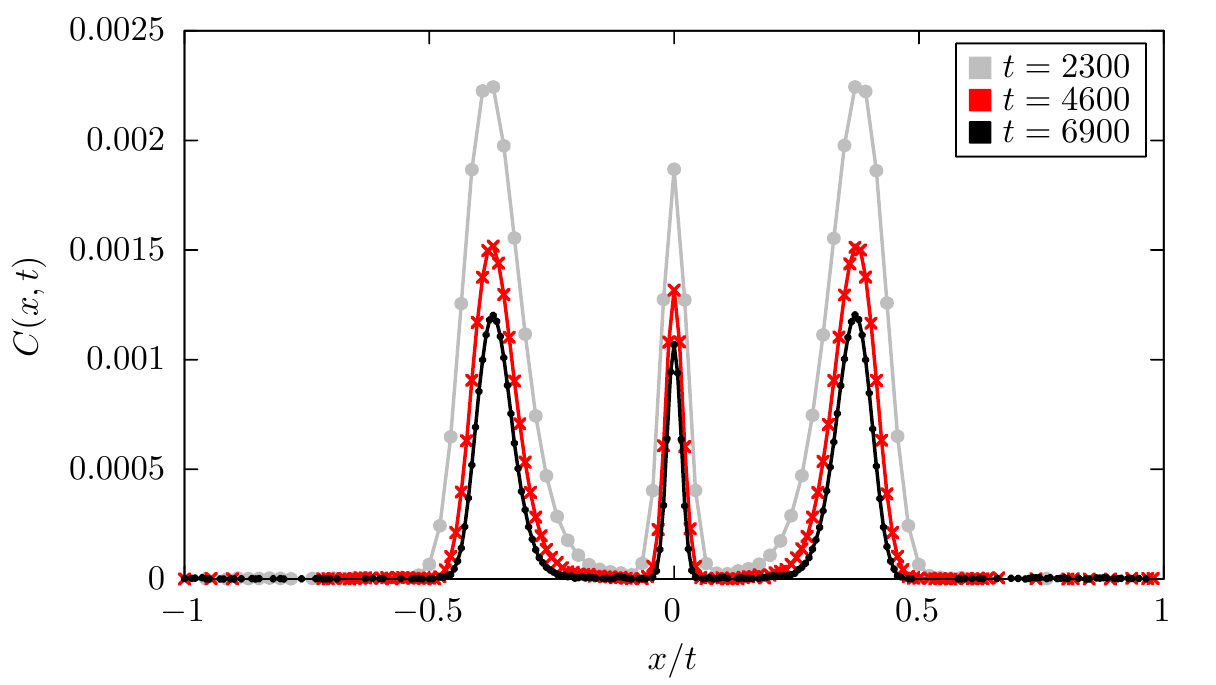}
  \caption{\label{fig:NonintegrableCorrelations} Dynamical correlation function~$C(x,t)$
  for the \emph{non-integrable} case and $\vec{p}=\vec{p}_{\infty}$. The correlation
  function exhibits three peaks: a central one and two side ones that move
  to the left and right.}
\end{figure}

We obtain the correlation function~$C(x,t)$ numerically by averaging over many
trajectories starting from random initial configurations that are sampled according
to the probability distribution~$\vec{p}$. The simplest stationary distribution~$\vec{p}$
is a \emph{maximum-entropy} (also \emph{infinite-temperature}) state~$\vec{p}_{\infty}$,
for which every configuration is equally likely. The numerically computed correlation
function profiles for~$\vec{p}=\vec{p}_{\infty}$ are shown in
Figures~\ref{fig:IntegrableCorrelations} and~\ref{fig:NonintegrableCorrelations}.
At small times both profiles exhibit a `heat' peak centered around position~$x=0$ and 
two `sound' peaks at the side that move to the left and to the right with fixed velocities. At longer
times, however, we observe that in the integrable case each one of the side peaks
splits into two separate contributions that move with very similar but different
velocities. In the \emph{non-integrable} model this does not appear to happen,
which is consistent with the finite number (3) of local conservation laws which equals the number of peaks.

\subsection{Hydrodynamic discrepancies on long time-scales}
To obtain more insight into the structure of~$C(x,t)$, let us consider the
hydrodynamic predictions for it. We focus on the \emph{non-integrable}
automaton, since in this case we expect that the MPS parametrization introduced
in Section~\ref{sec:GibbsStates} gives the complete set of stationary states.

We start by generalizing the correlation function~$C(x,t)$ to an arbitrary pair
of charges~$q_{\alpha}$, $q_{\beta}$, $\alpha,\beta\in\{1,2,3\}$,
\begin{eqnarray}
  C_{\alpha,\beta}(x,t)=
  \expval*{\left.q_{\alpha}\right._{x}^{t} \left.q_{\beta}\right._{0}}_{\vec{p}}
  -\expval*{\left.q_{\alpha}\right._{0}}_{\vec{p}}
  \expval*{\left.q_{\beta}\right._{0}}_{\vec{p}},
\end{eqnarray}
where~$\vec{p}=\vec{p}(\xi,\omega,\lambda)$ is a Gibbs state. Note that the
correlation function defined in~\eqref{eq:defCxt} is
a linear combination of~$C_{\alpha,\beta}(x,t)$,
\begin{eqnarray}
  C(x,t)=C_{2,2}(x,t)+C_{2,3}(x,t)+C_{3,2}(x,t)+C_{3,3}(x,t).
\end{eqnarray}

At the hydrodynamic scale in the linear order the correlation
function~$C_{\alpha,\beta}(x,t)$ obeys a linear homogeneous partial
differential equation (see e.g.\ the discussion
in~\cite[Sec.~2]{doyon2020lecturenotes} and~\cite{spohn2014nonlinear}),
\begin{eqnarray}
  \partial_t C_{\alpha,\beta}(x,t) +\partial_x \sum_{\gamma}
  A_{\alpha,\gamma} C_{\gamma,\beta}(x,t) = 0,
\end{eqnarray}
where~$A$ is a~$3\times 3$ matrix of derivatives of the expectation values of
currents with respect to the densities of charges,
\begin{eqnarray}
  A_{\alpha,\beta} = \frac{\partial \expval*{j_{\alpha\,0}}_{\vec{p}}}
  {\partial \expval{q_{\beta\,0}}_{\vec{p}}}.
\end{eqnarray}
Here the currents~$j_{\alpha}$ ($\alpha=1,2,3$) together with densities of
conserved charges obey the following continuity equation,
\begin{eqnarray} \label{eq:continuityEq}
  \left.q_{\alpha}\right._{2x}^{2t+2}-\left.q_{\alpha}\right._{2x}^{2t}
  + \left.j_{\alpha}\right._{2x+1}^{2t+1}-\left.j_{\alpha}\right._{2x-1}^{2t+1}=0.
\end{eqnarray}
Their explicit form is given in~\ref{app:currentsDetails}.

\begin{figure}
  \centering
  \includegraphics[width=0.78\textwidth]{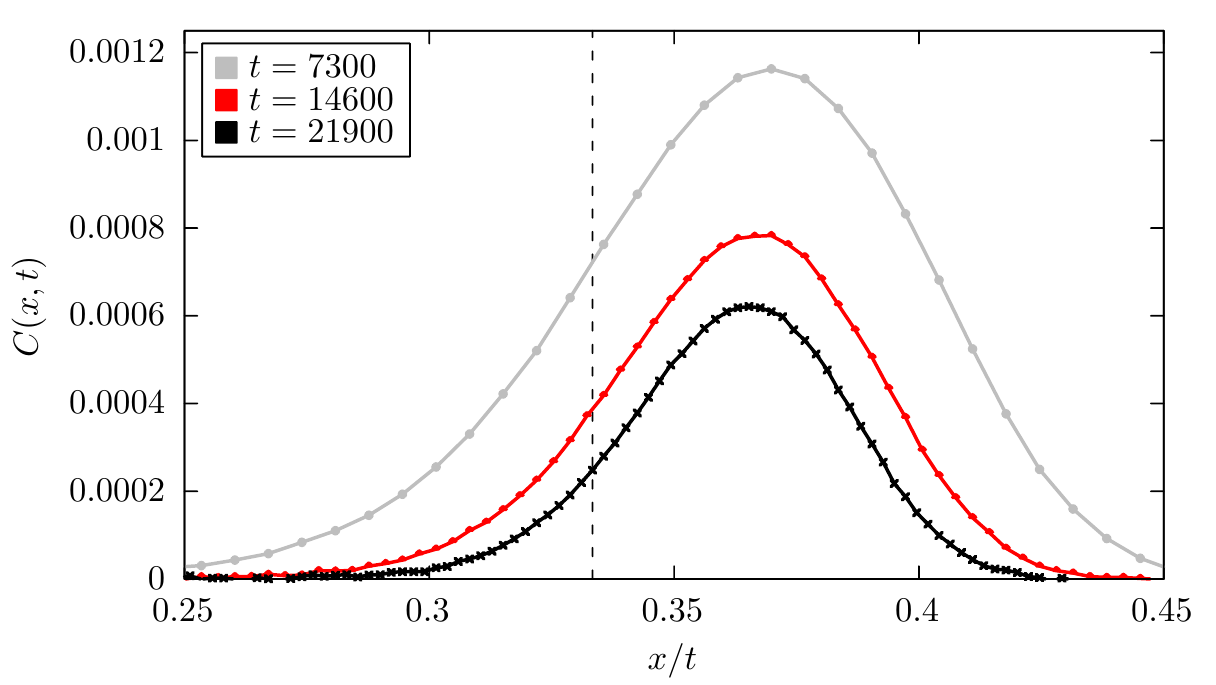}
  \caption{\label{fig:velocity1}
  Closer look at the right-moving peak of~$C(x,t)$ for the \emph{non-integrable}
  case. The dashed vertical line denotes the hydrodynamic prediction for the
  velocity of the peak~$v_{3}=\frac{1}{3}$.
  }
\end{figure}

At $t=0$, correlation function $C_{\alpha,\beta}(x,0)$ has a peak at~$x=0$  with
the tails decreasing exponentially in $|x|$. Therefore at time~$t$ in this regime
we expect the correlation function to (in general) exhibit three peaks at positions
$x=t v_{\alpha}$, where $v_{\alpha}$, $\alpha=1,2,3$ are the eigenvalues of~$A$. 
In the case of the maximum entropy state~$\vec{p}=\vec{p}_{\infty}$ the matrix~$A$
takes the following form (see~\ref{app:matADetails} for the details),
\begin{eqnarray} \label{eq:matAfinal}
  \left. A\right|_{\xi=\omega=\lambda=1} =
  \begin{bmatrix}
    0& \frac{1}{3} & \frac{1}{3} \\
    \frac{1}{6}& 0 & 0 \\
    \frac{1}{6}& 0 & 0 
  \end{bmatrix}=
  R
  \begin{bmatrix}
    -\frac{1}{3} & & \\
     &0& \\
     & &\frac{1}{3}
  \end{bmatrix}
  R^{-1},
\end{eqnarray}
for a suitably chosen similarity transformation~$R$. Therefore for a correlation
function between any linear combinations of densities of charges we expect the peaks to
be located at~$x=0$, and $x=\pm \frac{1}{3} t$. As we can see from the closeup of the
peak in Figure~\ref{fig:velocity1}, this does not seem to be the case: the velocity observed
in numerics is closer to $0.36$ than to $\frac{1}{3}$.

\begin{figure}
  \centering
  \includegraphics[width=0.78\textwidth]{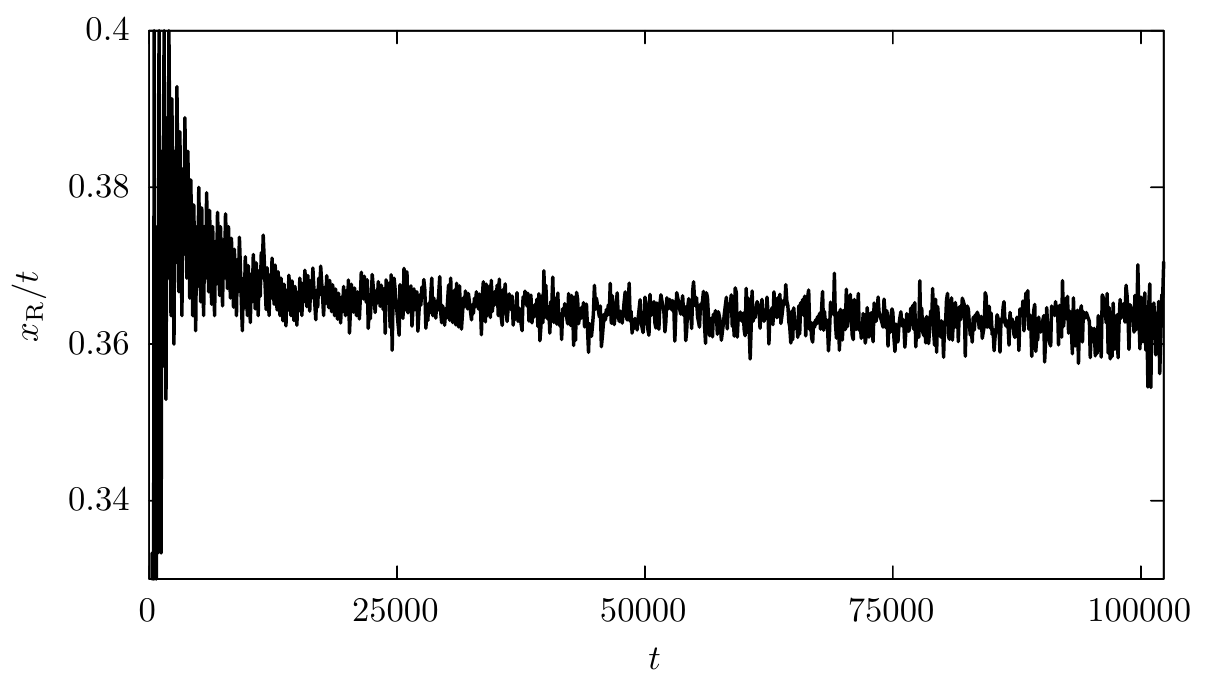}
  \caption{\label{fig:velocity}
  Perceived velocity of the right-moving peak of~$C(x,t)$ as a function of
  time. The velocity is determined as the position of the right-moving peak
  $x_{\mathrm{R}}$ divided by the time~$t$. Note that the hydrodynamic
  prediction for the velocity is~$v_{\mathrm{R}}=\frac{1}{3}$, which is not
  matched by the numerical results on very long timescales. However, a small
  drift in the velocity can be observed (see Figure~\ref{fig:tempFigure}).
  }
\end{figure}

\begin{figure}
  \centering
  \includegraphics[width=0.78\textwidth]{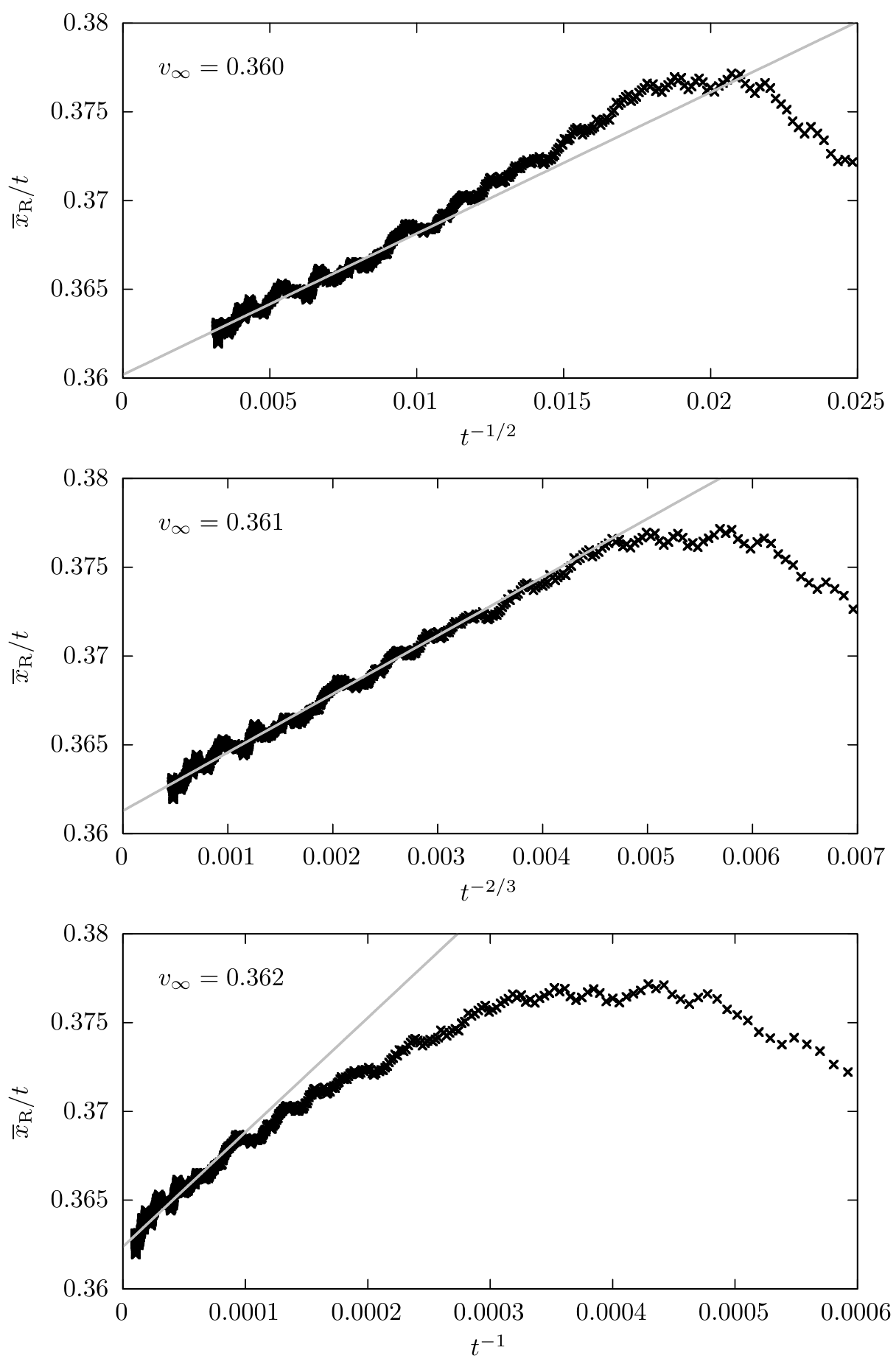}
  \caption{\label{fig:tempFigure}
    Perceived velocity of the right-moving peak of~$C(x,t)$ as a function of
    different choices of the algebraic scaling of time $t^{-\gamma}$. Here
    instead of the position $x_{\mathrm{R}}$ we report the moving-time-averaged
    position
    $\overline{x}_{\mathrm{R}}=\frac{1}{\Delta}\int_{t-\Delta/2}^{t+\Delta/2}
    x_{\mathrm{R}}(t)$, with $\Delta=5000$.  The grey solid line corresponds to the best
    fit of the linear function and for each $\gamma$ we extract the limit
    $v_{\infty}=\lim_{t\to\infty} \overline{x}_{\mathrm{R}}/t$ from the fit.
    We do not have any analytical prediction for the scaling exponent $\gamma$,
    however the numerical fits reported above suggest $\gamma$ takes a value
    inside the interval $(1/2,1)$, and it is close to $2/3$. For all the
    reported cases the extracted asymptotic velocity $v_{\infty}$ disagrees
    with the hydrodynamic prediction ($v_3=\frac{1}{3}$).
    }
\end{figure}

To investigate this issue further, in Figure~\ref{fig:velocity} we plot the ratio
between the position of the right peak $x_{\mathrm{R}}$ and time~$t$ on longer
time-scales. The numerically perceived velocity shows a very slow, almost insignificant drift away from~$0.36$.
In Figure~\ref{fig:tempFigure} we plot the averaged values of velocities with respect
to different choices of the scaling functions of time, $t^{-\gamma}$. The numerical data
is well approximated by
\be
\frac{\overline{x}_{\mathrm{R}}}{t}-v_{\infty}(\gamma)\propto t^{-\gamma},
\label{scaling}
\ee
for appropriately determined $v_{\infty}(\gamma)$ (extracted from the fit) and
$\gamma$.  The quality of scaling is not very sensitive to varying $\gamma$
between $1/2$ and $1$, but perhaps the best scaling is seen for $\gamma=2/3$
where the asymptotic velocity reads $v_{\infty} = 0.361$.  Interestingly, the
asymptotic velocity $v_{\infty}(\gamma)$ does \emph{not} match the hydrodynamic
prediction $v_3=1/3$. Even though we cannot exclude the possibility of $x/t$
reaching $v_3$ by eventually breaking the scaling \eqref{scaling}, the
time-scales associated with this convergence should be extremely long (note
that the longest times in our data are of the order $10^5$). This suggests the
existence of an additional \emph{quasi-local conserved quantity} or an \emph{almost
conserved local quantity} in this dynamical system. The existence of such
quantities in non-integrable locally interacting systems should be of general
interest beyond the context of RCA.

\section{Conclusions}
We have proposed two reversible cellular automata, which generalise the
interacting particle-like dynamics of Rule 54~\cite{bobenko1993two,RCA54review}
to two particle species. In addition to simple one-species solitons that behave
analogously to solitons in RCA54, the models exhibit more complicated
multi-colour particles, which are, however, not stable under scattering.
Empirical evidence suggests that one of the automata is an integrable system,
while the other one possesses only three local integrals of motion.  These
three conservation laws are common to both models, therefore a class of
generalised Gibbs states with nonzero chemical potentials corresponding to
these observables is stationary for both automata. For periodic boundaries we
find a staggered matrix-product formulation of this class of states, consisting
of matrices that obey a cubic algebraic relation analogous
to~\cite{prosen2017exact}.  We extend this class to four-parameter stationary
states of the integrable automaton, which we express both in a
patch-state-ansatz (analogous to~\cite{prosen2016integrability}) and
matrix-product-state form. We numerically study the behaviour of
spatio-temporal correlation functions in the non-integrable model. We find
significant deviations from the hydrodynamic prediction obtained by assuming
local thermalization in the presence of no other conservation laws apart from
the three known ones.

The collection of results presented here leaves many open questions. First, it
would be interesting to find Yang-Baxter formulation for the integrable model
and thus rigorously establish the integrability of the first automaton. Perhaps
recent ideas of~\cite{pozsgay2021yang,gombor2021integrable} could provide a
good starting point.  Moreover, it is still not understood which aspects of
solvability of Rule 54 carry on to these models. For instance, it would be
interesting to find non-equilibrium stationary states of the boundary driven
setup (analogous
to~\cite{prosen2016integrability,prosen2017exact,inoue2018two}), explore how
(non)integrability affects the form of stationary states, and whether the exact
treatment of large-deviation statistics is feasible~\cite{buca2019exact}.
    From a physical perspective, it would be interesting to understand 
    the nature of transport in these models. In particular, preliminary analysis
    suggests anomalous broadening of sound peaks, while heat peaks exhibit
    diffusive spreading, but a more extensive numerical exploration remains to be
    done.

Probably the most urgent question concerns the explanation of the disagreement
between the behaviour of correlation functions and the hydrodynamic prediction.
Even though the numerics can not rule out the eventual relaxation to the
hydrodynamic values, the time-scales for that are surprisingly long and one
should understand what the mechanism for this could be (see e.g.
Refs.~\cite{lebowitz2018ballistic,dicintio2018transport,dhar2019transport,cao2018incomplete}
for some recent related results). Precisely because of mathematical and
conceptual simplicity of the model we hope that further analytic progress will
be possible in future.

\section*{Acknowledgements}
We thank M.\ Medenjak and V.\ Popkov for useful discussions.  This work has
been supported by the European Research Council (ERC) under the Advanced Grant
No.\ 694544 -- OMNES, by the Slovenian Research Agency (ARRS) under the Program
P1-0402 and by the Engineering and Physical Sciences Research Council (EPSRC)
through the grant EP/S020527/1.

\appendix
\section{Conserved quantities}\label{app:consQuant}
\subsection{Algorithm}
The numerical procedure to search for local conserved quantities is a variant of the algorithm
introduced in Sec.~7 of~\cite{prosen2007chaos}, which can be understood as an exact
diagonalization of the time-evolution operator projected to the space of extensive local
observables. 

We start by introducing a convenient orthogonal basis of local observables,
\be
\oo{\hat{0}} = \oo{0}+\oo{1}+\oo{2},\qquad
\oo{\hat{1}} = \oo{1}-\oo{2},\qquad
\oo{\hat{2}} = -2\oo{0}+\oo{1}+\oo{2}, 
\ee
where $\oo{\hat{0}}$ is the identity observable (i.e.\ its expectation value is $1$ for each
state) and it has the role of the unit element in the commutative algebra of local observables,
while $\oo{\hat{1}}$ and $\oo{\hat{2}}$ are two other independent basis elements chosen so that
the basis is orthogonal.\footnote{The inner product that is used here is the expectation value of
the product of local observables in the maximum entropy state
$a,b \mapsto \expval{a b}_{\vec{p}_{\infty}}$, with
$\vec{p}_{\infty}\propto\left[1\ 1\ \ldots 1\right]^T$.}
Note that by definition, the string of $\hat{0}$ in the definition of an observable with
a larger support can be removed and replaced by a translation operator,
\be
\oo{\underbrace{\hat{0}\hat{0}\ldots \hat{0}}_{m} \hat{s}_1\hat{s}_2\cdots \hat{s}_k}
=\oo{\hat{s}_1\hat{s}_2\cdots \hat{s}_k}_{m+1},
=\eta_m(\oo{\hat{s}_1\hat{s}_2\cdots \hat{s}_k}),
\ee
which follows from
\be
a \cdot \oo{\hat{0}}_x=\oo{\hat{0}}_x \cdot a = a,
\ee
for any local observable $a$.

We denote the local time-evolution
operator written in this basis by $\hat{U}^{\alpha}_{x-1 x x+1}$ and it is given by 
\be
\hat{U}^{\alpha}_{1 2 3} = \left(R\otimes R\otimes R\right)^{-1} U^{\alpha}_{1 2 3}
\left(R\otimes R\otimes R\right),
\ee
with $R$ being the local basis transformation,
\be
R=
\begin{bmatrix}
    1&0&-2\\ 1&1&1\\ 1&-1&1
\end{bmatrix}.
\ee
We proceed by defining the following projector to local observables with support $r$,
\be
\hat{P}_r
\left.
\oo{\underbrace{\hat{0}\hat{0}\ldots\hat{0}}_{2m}\hat{s}_1\hat{s}_2 \ldots\hat{s}_{k}}
\right|_{s_1+s_2>0,s_k>0}
=
\begin{cases}
    \oo{\hat{s}_1 \hat{s}_2 \ldots \hat{s}_k},& k\le r,\\
    0,&k>r,
\end{cases},
\ee
which maps all local observables with support $k<r$ to an equivalent 
observable that acts nontrivally on the section of the chain $[1,...,k]$.
The important feature of this map is that it provides a unique way to express any extensive
observable with (quasi)local density:
\be \label{eq:nonUniqueness}
A=\sum_{j=1}^n \eta_{2j}(a) = 
\lim_{r\to\infty} \sum_{j=1}^n \eta_{2j}(\hat{P}_r a),
\ee
where the left-hand side is not unique (i.e.\ there exist many different equivalent choices
of $a$ that give the same $A$), while the expression on the r.h.s.\ is.
Exploiting this uniqueness, we can now interpret the local densities of extensive
conserved charges with support smaller than $r$ to be the eigenvectors corresponding
to the eigenvalue $1$ of the reduced time-evolution operator $\hat{U}_r$ defined as, 
\be \label{eq:redTEop} \fl
\hat{U}_{r}^{\alpha} = \hat{P}_{r}
\hat{U}^{\alpha}_{123}
\hat{U}^{\alpha}_{345}\cdots
\hat{U}^{\alpha}_{k-2,k-1,k}
\eta_1 \hat{P}_r,\qquad
k=
\begin{cases}
    r+1,& r\equiv 0\pmod{2},\\
    r+2,& r\equiv 1\pmod{2},
\end{cases}
\ee
where $k=k(r)$ is either $r+1$ or $r+2$ (depending on the parity of $r$). Therefore, to find
the number of local conservation laws with small support $r$ one can just diagonalize the 
above operator, which is how Table~\ref{tab:numCQs} was obtained.

\subsection{Conservation law with support 4}
Diagonalising the reduced time-evolution operator~\eqref{eq:redTEop} one can also find 
the explicit form of conservation laws, and as a special example we can obtain the first
higher conservation law that is conserved in the first automaton $\chi_1$, but not for $\chi_2$.
Note, however, that not only is the local density not unique (as noted in
Eq.~\eqref{eq:nonUniqueness}), but also any linear combination of local conservation laws with
support smaller than $r$ is also conserved and there are many different choices
of a set of linearly independent conserved charges (unless we require
orthogonality with respect to some inner product). One such choice of the new conservation law
$Q_4$,
\be
Q_4=\sum_{j=1}^n \eta_{2j}(\tilde{q}_4),
\ee
is determined by the following local density
\be \fl
\eqalign{
    \tilde{q}_4 &=
\frac{1}{6}\oo{\hat{0}\hat{1}\hat{0}\hat{1}}
+ \frac{1}{18}\oo{\hat{0}\hat{1}\hat{1}\hat{0}}
+ \frac{1}{36}\oo{\hat{0}\hat{1}\hat{1}\hat{2}}
- \frac{1}{12}\oo{\hat{0}\hat{1}\hat{2}\hat{1}}
+ \frac{1}{54}\oo{\hat{0}\hat{2}\hat{0}\hat{2}}
+ \frac{1}{12}\oo{\hat{0}\hat{2}\hat{1}\hat{1}}
- \frac{1}{54}\oo{\hat{0}\hat{2}\hat{2}\hat{0}}\\
&+ \frac{1}{108}\oo{\hat{0}\hat{2}\hat{2}\hat{2}}
+ \frac{1}{6}\oo{\hat{1}\hat{0}\hat{1}\hat{0}}
+ \frac{1}{6}\oo{\hat{1}\hat{1}\hat{0}\hat{0}}
- \frac{1}{8}\oo{\hat{1}\hat{1}\hat{1}\hat{1}}
+ \frac{1}{12}\oo{\hat{1}\hat{1}\hat{2}\hat{0}}
- \frac{1}{24}\oo{\hat{1}\hat{1}\hat{2}\hat{2}}
- \frac{1}{12}\oo{\hat{1}\hat{2}\hat{1}\hat{0}}\\
&+ \frac{1}{24}\oo{\hat{1}\hat{2}\hat{1}\hat{2}}
+ \frac{1}{8}\oo{\hat{1}\hat{2}\hat{2}\hat{1}}
+ \frac{1}{54}\oo{\hat{2}\hat{0}\hat{2}\hat{0}}
+ \frac{1}{36}\oo{\hat{2}\hat{1}\hat{1}\hat{0}}
+ \frac{1}{72}\oo{\hat{2}\hat{1}\hat{1}\hat{2}}
+ \frac{1}{24}\oo{\hat{2}\hat{1}\hat{2}\hat{1}}
- \frac{1}{54}\oo{\hat{2}\hat{2}\hat{0}\hat{0}}\\
&- \frac{1}{24}\oo{\hat{2}\hat{2}\hat{1}\hat{1}}
+ \frac{1}{108}\oo{\hat{2}\hat{2}\hat{2}\hat{0}}
- \frac{1}{72}\oo{\hat{2}\hat{2}\hat{2}\hat{2}}.
}
\ee 
This choice of the linear combination of the first $4$ conservation laws was chosen
so that the state introduced in Section~\ref{sec:statesHigherCQs} satisfies
\be
\vec{p} \propto \xi^{Q_1} \omega^{Q_2} \lambda^{Q_3} \mu^{Q_4}.
\ee
To see that this really holds, we note that by definition the extensive observable given
by the density $q_4$ defined in terms of patch-state-ansatz values $t_{s_1 s_2 s_3 s_4}$ 
(see Subsection~\ref{sec:PSA}) as,
\be
q_{4}=\sum_{s_1,s_2,s_3,s_4\in\{0,1,2\}}
\left.\frac{\partial t_{s_1 s_2 s_3 s_4}}{\partial \mu}\right|_{\xi,\omega,\lambda,\mu\to1}
\oo{s_1 s_2 s_3 s_4},
\ee
is conserved. Using this definition, we obtain
\be \fl
\eqalign{
    q_4&= \oo{0100}+ \oo{0101}+ \oo{0111}+ \oo{0122}+
    \oo{0200}+ \oo{0202}+ \oo{0211}+ \oo{0222}+ \oo{1000}\\
    &\, +\oo{1001}+ \oo{1010}+ \oo{1011}+ \oo{1012}+
    \oo{1022}+ \oo{1111}+ \oo{1200}+ \oo{1202}+ \oo{1211}\\&\,+
    \oo{1222}+ \oo{2000}+ \oo{2002}+ \oo{2011}+ \oo{2020}+
    \oo{2021}+ \oo{2022}+ \oo{2100}+ \oo{2101}\\&\,+
    \oo{2111}+ \oo{2122}+ \oo{2222}.
}
\ee 
To prove that $\tilde{q}_4$ and $q_4$ give the same extensive observable $Q_4$
(even though $\tilde{q}_4\neq q_4$), one just needs to verify that following
holds,
\be
\hat{P}q_4=\hat{P}_4\tilde{q}_4=\tilde{q}_4,
\ee 
which can be checked explicitly.

\section{Asymptotic probabilities in Gibbs ensembles}\label{app:gibbsHDL}
Expectation value of a basis observable $\oo{s_1 s_2\ldots s_{2m}}$ with
support~$2m$ in the Gibbs state~$\vec{p}$ is expressed as the following
product of matrices,
\begin{eqnarray}
  \expval*{\oo{s_1 s_2\ldots s_{2m}}}_{\vec{p}}=
  \frac{1}{\tr T^{n}} \tr\left(\W_{s_1}\V_{s_2}\cdots \V_{s_{2m}} T^{n-m}\right),
\end{eqnarray}
where we introduced the \emph{transfer matrix} $T$, defined
as the sums of products of matrices~$\W_{s}$, $\V_{s}$ on consecutive sites,
\begin{eqnarray}\fl
  T\mkern-5mu=\mkern-5mu\sum_{s_1,s_2}\W_{s_1}\V_{s_2}\mkern-5mu=\mkern-5mu
  \sum_{s_1,s_2}\V_{s_1}\W_{s_2}\mkern-5mu=\mkern-5mu
  \begin{bmatrix}
    3 & \lambda + \xi + \xi^{-1} & \omega + \xi + \xi^{-1} \\
    \omega + \xi+ \xi^{-1} & 2+\lambda\omega & 1+ \omega\left(\xi+\xi^{-1}\right)\\
    \lambda + \xi + \xi^{-1} & 1+ \lambda\left(\xi+\xi^{-1}\right) & 2+\lambda\omega
  \end{bmatrix}\mkern-4mu.
\end{eqnarray}
When the system size is much bigger than the support of the observable in question,
the expression simplifies into
\begin{eqnarray}
  \lim_{n\to\infty} \expval*{\oo{s_1 s_2\ldots s_{2m}}}_{\vec{p}}=
  \frac{\Lambda^{-m}}{\braket{l}{r}}\bra{l}\W_{s_1}\V_{s_2}\cdots \V_{s_{2m}}\ket{r},
\end{eqnarray}
where~$\Lambda$, $\ket{r}$ and $\bra{l}$ are the leading eigenvalue and the
corresponding right and left eigenvectors of the transfer matrix~$T$,
\begin{eqnarray}
  T \ket{r} = \Lambda \ket{r},\qquad \bra{l} T = \Lambda \bra{l}.
\end{eqnarray}
Equivalently, for the other time-step parity the expectation value can be
expressed by simply exchanging $\W_s\leftrightarrow\V_s$,
\begin{eqnarray}
  \lim_{n\to\infty} \expval*{\oo{s_1 s_2\ldots s_{2m}}}_{\vec{p}^{\prime}}=
  \frac{\Lambda^{-m}}{\braket{l}{r}}\bra{l}\V_{s_1}\W_{s_2}\cdots \W_{s_{2m}}\ket{r}.
\end{eqnarray}

\section{Cubic algebra for higher stationary states} \label{sec:matXY}
The components of~$\vX$, $\vY$ that together with $\vW$ and $\vV$ satisfy Eqs.~\eqref{eq:WWW}
and~\eqref{eq:WWWcompt} take the following form,
\be \fl
\X_0 = 
\begin{bmatrix}
    1 & 0 & 0 & 0 & 1 & 0 & 1 \\
    0 & 1 & 0 & 1 & 0 & 0 & 0 \\
    0 & 0 & 1 & 0 & 0 & 1 & 0 \\
    0 & 0 & 0 & 0 & 0 & 0 & 0 \\
    0 & 0 & 0 & 0 & 0 & 0 & 0 \\
    0 & 0 & 0 & 0 & 0 & 0 & 0 \\
    0 & 0 & 0 & 0 & 0 & 0 & 0
\end{bmatrix},\qquad
&\Y_0 = 
\begin{bmatrix}
    1 & 1 & 1 & 0 & 0 & 0 & 0 \\
    0 & 0 & 0 & 0 & 0 & 0 & 0 \\
    0 & 0 & 0 & 0 & 0 & 0 & 0 \\
    \mu  & \mu  & 1 & 0 & 0 & 0 & 0 \\
    0 & 0 & 0 & 0 & 0 & 0 & 0 \\
    \mu  & 1 & \mu  & 0 & 0 & 0 & 0 \\
    0 & 0 & 0 & 0 & 0 & 0 & 0
\end{bmatrix},\\
\fl
\X_1=
\begin{bmatrix}
    0 & 0 & 0 & 0 & 0 & 0 & 0 \\
    0 & 0 & 0 & 0 & 0 & 0 & 0 \\
    0 & 0 & 0 & 0 & 0 & 0 & 0 \\
    0 & 0 & \xi^2 & 1 & \mu  & \xi\omega & 0 \\
    \xi^2 & \xi^2 & 0 & 0 & 0 & 0 & \xi\omega \\
    0 & 0 & 0 & 0 & 0 & 0 & 0 \\
    0 & 0 & 0 & 0 & 0 & 0 & 0 
\end{bmatrix},\qquad
&\Y_1 = 
\begin{bmatrix}
    0 & 0 & 0 & 0 & 0 & 0 & 0 \\
    0 & 0 & \frac{\omega}{\xi} & \frac{1}{\xi^2}  & \frac{\mu}{\xi^2}  & 0 & 0 \\
    0 & 0 & 0 & 0 & 0 & 0 & 0 \\
    0 & 0 & 0 & 0 & 0 & 0 & 0 \\
    0 & 0 & \xi\omega & 1 & 1 & 0 & 0 \\
    0 & 0 & 0 & 0 & 0 & 0 & 0 \\
    0 & 0 & 0 & 0 & 0 & 0 & 0
\end{bmatrix},
\ee
and
\be \fl
\X_2 = 
\begin{bmatrix}
    0 & 0 & 0 & 0 & 0 & 0 & 0 \\
    0 & 0 & 0 & 0 & 0 & 0 & 0 \\
    0 & 0 & 0 & 0 & 0 & 0 & 0 \\
    0 & 0 & 0 & 0 & 0 & 0 & 0 \\
    0 & 0 & 0 & 0 & 0 & 0 & 0 \\
    0 & \xi^2 & 0 & \lambda\xi & 0 & 1 & \mu  \\
    \xi^2 & 0 & \xi^2 & 0 & \lambda\xi & 0 & 0 
\end{bmatrix},\qquad
\Y_2 =
\begin{bmatrix}
    0 & 0 & 0 & 0 & 0 & 0 & 0 \\
    0 & 0 & 0 & 0 & 0 & 0 & 0 \\
    0 & \frac{\lambda}{\xi} & 0 & 0 & 0 & \frac{1}{\xi^2} & \frac{\mu}{\xi^2} \\
    0 & 0 & 0 & 0 & 0 & 0 & 0 \\
    0 & 0 & 0 & 0 & 0 & 0 & 0 \\
    0 & 0 & 0 & 0 & 0 & 0 & 0 \\
    0 & \lambda\xi & 0 & 0 & 0 & 1 & 1 \\
\end{bmatrix}.
\ee

\section{Euler scale hydrodynamics}\label{app:hydroDetails}
\subsection{Currents}\label{app:currentsDetails}
The local density of the currents that satisfy the continuity equation~\eqref{eq:continuityEq}
is
\begin{eqnarray}
  \eqalign{
    \left.j_1\right._{x} = \sum_{s=0}^{2}
    \left(
    \oo{sss}_x - \oo{s0s}_x
    -\oo{ss}_x-\oo{ss}_{x+1}
    \right)\\
    \left.j_2\right._{x} = \oo{112}_x-\oo{122}_x,\qquad
    \left.j_3\right._{x} = \oo{221}_x-\oo{211}_x.
  }
\end{eqnarray}

\subsection{Matrices of cross correlations}\label{app:matADetails}
To express $A$, it is convenient to use the chain rule,
\begin{eqnarray} \label{eq:matAchainrule}
  A=L K^{-1},
\end{eqnarray}
where $K$ and $L$ are $3\times 3$ matrices with the matrix elements given by,
\begin{eqnarray}\fl
  K_{\alpha,\beta} = \frac{\partial \expval*{q_{\alpha\,0}}_{\vec{p}}}
  {\partial \mu_{\beta}},\qquad
  L_{\alpha,\beta} = \frac{\partial \expval*{j_{\alpha\,0}}_{\vec{p}}}
  {\partial \mu_{\beta}},\qquad
  (\mu_1,\mu_2,\mu_3)=(\log\xi,\log\omega,\log\lambda).
\end{eqnarray}
Here~$\mu_{\alpha}$, $\alpha\in\{1,2,3\}$ are chemical potentials associated to
the three conserved charges. A convenient way to evaluate the relevant expectation values
is to use the matrix product from~\ref{app:gibbsHDL}, which in
the limit $\xi,\omega,\lambda\to 1$ gives the following form of $K$ and $L$,
\be
\left.K\right|_{\xi=\omega=\lambda=1} = 
\begin{bmatrix}
    \frac{4}{9} & 0 & 0 \\
    0 & \frac{4}{27} & \frac{2}{27} \\
    0 & \frac{2}{27} & \frac{4}{27}
\end{bmatrix},\qquad
\left.L\right|_{\xi=\omega=\lambda=1} =
\begin{bmatrix}
    0 & \frac{1}{3} & \frac{1}{3} \\
    \frac{1}{6} & 0 &  0 \\
    \frac{1}{6} & 0 &  0
\end{bmatrix}.
\ee
Plugging these into~\eqref{eq:matAchainrule}, we obtain precisely $A$ as given
in~\eqref{eq:matAfinal}.

\section*{References}
\bibliographystyle{iopart-num}
\bibliography{bibfile}
\end{document}